\documentclass[12pt,english]{article}
\usepackage[T1]{fontenc}
\usepackage[latin1]{inputenc}
\usepackage{geometry}
\usepackage{caption}
\DeclareCaptionFormat{capname}{\textbf{ #1.} #3}
\DeclareCaptionLabelFormat{Figurestyle}{\textbf{FIG. #2}}

\usepackage{setspace}
\makeatletter

\usepackage{epsfig}
\usepackage{amssymb}
\usepackage{graphicx}
\usepackage[pdftex,dvipsnames,usenames]{color}

\usepackage{babel}
\makeatother
\begin{document}
\begin{doublespace}

\title{Self-similar shear-thickening behavior in CTAB/NaSal surfactant solutions}
\end{doublespace}
\date{}
\maketitle

\author{\begin{center} Mukund Vasudevan
\\ {\it{Department of Energy, Environmental and Chemical Engineering, Washington University in St.Louis, St.Louis MO 63130}}\\

Amy Shen
\\ {\it{Department of Mechanical, Aerospace and Structural Engineering, Washington University in St.Louis, St.Louis MO 63130}}\\

Bamin Khomami
\\ {\it{Department of Chemical and Biomolecular Engineering, University of Tennessee, Knoxville, TN 37919}}

Radhakrishna Sureshkumar\footnote{Electronic mail: suresh@wustl.edu}
\\ {\it{Center for Materials Innovation, and Department of Energy, Environmental and Chemical Engineering, Washington University in St.Louis, St.Louis MO 63130}}\\\end{center}
}

 \newcommand{\Amin} {Amin {\it {et al.}} (2001)}
 \newcommand{\Appell} {Appell {\it {et al.}} (1982)} \newcommand{\Arora} {Arora {\it{et al.}} (2002)} \newcommand{\Aswal} {Aswal {\it{et al.}} (1998)} \newcommand{\Azzouzi} {Azzouzi {\it{et al.}} (2005)} \newcommand{\Bando} {Bandyopadhyay {\it {et al.}} (2000)} \newcommand{\Bandob} {Bandyopadhyay and Sood (2001)} \newcommand{\Callaghan} {Britton and Callaghan (1999)} \newcommand{\Callaghanb} {Britton and Callaghan (1997)} \newcommand{\Cappelaere} {Cappelaere {\it {et al.}} (1994)} \newcommand{\Cappelaereb} {Cappelaere and Cressely (2000)} \newcommand{\Cates} {Cates and Candau (1990)} \newcommand{\Catesb} {Cates (1996)} \newcommand{\Catesc} {Cates and Candau (2001)} \newcommand{\Catesd} {Cates (1987a)} \newcommand{\Catese} {Cates (1987)} \newcommand{\Catesf} {Cates (1988)} \newcommand{\Clausen} {Clausen {\it{et al.}} (1992)} \newcommand{\Cressely} {Cressely and Hartmann (1998a)} \newcommand{\Cummins} {Cummins {\it {et al.}} (1987)} \newcommand{\Doi} {Doi and Edwards (1986)} \newcommand{\Fielding} {Fielding and Olmsted (2003)} \newcommand{\Fieldingb} {Fielding and Olmsted (2004)} \newcommand{\Fieldingc} {Fielding and Olmsted (2005)} \newcommand{\Fischer} {Fischer (2000)} \newcommand{\Fujio} {Fujio (1998)} \newcommand{\Garg} {Garg {\it {et al.}} (2006)} \newcommand{\Gelbart} {Gelbart {\it {et al.}} (1994)} \newcommand{\Goyala} {Goyal {\it {et al.}} (1989)} \newcommand{\Goyalb} {Goyal {\it {et al.}} (1991)} \newcommand{\Goyalc} {Goyal {\it {et al.}} (1992)} \newcommand{\Hamley} {Hamley (2000)} \newcommand{\Hartmann} {Hartmann and Cressely (1997)} \newcommand{\Hartmannb} {Hartmann and Cressely (1998b)} \newcommand{\Hartmannc} {Hartmann and Cressely (1997)} \newcommand{\Hu} {Hu {\it {et al.}} (1998)} \newcommand{\Hub} {Hu and Mathys (1995)} \newcommand{\Huc} {Hu and Mathys (1996)} \newcommand{\Imae} {Imae and Ikeda (1986)}
 \newcommand{\Imaeb} {Imae {\it {et al.}} (1985)} \newcommand{\Imaec} {Imae and Kohsaka (1992)} \newcommand{\Jana} {Jana {\it {et al.}} (2004)} \newcommand{\Kim} {Kim and Yang (2000)} \newcommand{\Kumar} {Eggert and Kumar (2004)} \newcommand{\Kumarb} {Shankar and Kumar (2004)} \newcommand{\Kumaran} {Kumaran and Muralikrishnan (2000)} \newcommand{\Kumaranb} {Kumaran {\it {et al.}} (1994)} \newcommand{\Kumaranc} {Muralikrishnan and Kumaran (2002)} \newcommand{\Lina} {Lin {\it {et al.}} (1992)} \newcommand{\Linb} {Lin {\it {et al.}} (1994)} \newcommand{\Lin} {Lin {\it {et al.}} (2001)} \newcommand{\Liu} {Liu and Pine (1996)} \newcommand{\MacKintosh} {MacKintosh (1990)} \newcommand{\Mohan} {Mohan {\it {et al.}} (1976)} \newcommand{\Myska} {Myska and Mik (2004)} \newcommand{\Macias} {Mac{$\acute{i}$}as {\it {et al.}} (2003)} \newcommand{\Mendez} {Mend{$\acute{e}$}z-S{$\acute{a}$}nchez {\it {et al.}} (2003)} \newcommand{\Nowak} {Nowak (1998)} \newcommand{\Nowakb} {Nowak (2001)} \newcommand{\Oelschlaeger} {Oelschlaeger {\it {et al.}} (2002)} \newcommand{\Ouchi} {Ouchi {\it {et al.}} (2006)} \newcommand{\Penfold} {Penfold {\it {et al.}} (1991)} \newcommand{\Protzl} {Pr{$\ddot{o}$}tzl and Springer (1997)} \newcommand{\Russel} {Russell {\it{et al.}} (1989)} {\newcommand{\Rao} {Rao {\it{et al.}} (1987)} \newcommand{\Rehage} {Rehage and Hoffmann (1991)} \newcommand{\Rothstein} {Rothstein (2003)} \newcommand{\Shankar} {Shankar and Kumar (2004)} \newcommand{\Shankarb} {Eggert and Kumar (2004)} \newcommand{\Shen} {Shen {\it {et al.}} (2007)} \newcommand{\Shikata} {Shikata {\it {et al.}} (2002)} \newcommand{\Shikatab} {Shikata {\it {et al.}} (1987)} \newcommand{\Shikatac} {Shikata {\it {et al.}} (1988)} \newcommand{\Shikatad} {Shikata {\it {et al.}} (1994)} \newcommand{\Shukla} {Shukla {\it {et al.}} (2006)} \newcommand{\Tardos} {Tardos (1984)} \newcommand{\Tsujii} {Tsujii (1998)} \newcommand{\Wheeler} {Wheeler {\it {et al.}} (1998)} \newcommand{\Wunderlich} {Wunderlich {\it {et al.}} (1987)} \newcommand{\Zana} {Zana (1987)}
\newcommand{\cds} {$C_{D}$} \newcommand{\csd} {$C_{s}$} \newcommand{\etal} {\it {et al.}} \newcommand{\lp} {$l_{p}$} \newcommand{\appvisc} {$\eta_{app}$} \newcommand{\zsv} {$\eta_{0}$} \newcommand{\csr} {$\dot{\gamma}_{c}$} \newcommand{\nsc} {$\psi_{1}$} \newcommand{\nsco} {$\psi_{1,0}$} \newcommand{\nscm} {$\psi_{1,max}$} \newcommand{\sr} {$\dot{\gamma}$} \newcommand{\R} {\emph{R}} \newcommand{\Rey} {\textbf{{Re}}} \newcommand{\G} {\textbf{\emph{G}}} \newcommand{\relgel} {$\lambda_{gel}$}
\newcommand{\Bullethuge} {\hbox to 13pt{\hss\huge $\bullet$}}
\newcommand{\circhuge} {\hbox to 13pt{\hss\huge $\circ$}}

\begin{abstract}
The effect of salt concentration {\csd} on the critical shear rate $\dot{{\gamma}_{c}}$ required for the onset of shear thickening and apparent relaxation time $\lambda$ of the shear-thickened phase, has been investigated systematically for dilute CTAB/NaSal solutions. Experimental data suggest a self-similar behavior of {\csr} and $\lambda$ as functions of {\csd}. Specifically, $\dot{{\gamma}_{c}}$ $\sim$ ${C_{s}}^{-6}$ whereas $\lambda$ $\sim$ ${C_{s}}^{6}$ such that an effective Weissenberg number \emph{We} $\equiv$  $\lambda \dot{{\gamma}}$ for the onset of the shear thickened phase is only weakly dependent on $C_{s}$. A procedure has been developed to collapse the apparent shear viscosity versus shear rate data obtained for various values of $C_{s}$ into a single master curve. The effect of $C_{s}$ on the elastic modulus and mesh size of the shear-induced gel phase for different surfactant concentrations is discussed. Experiments performed using different flow cells (Couette and cone-and-plate) show that {\csr}, $\lambda$ and the maximum viscosity attained are geometry-independent. The elastic modulus of the gel phase inferred indirectly by employing simplified hydrodynamic instability analysis of a sheared gel-fluid interface is in qualitative agreement with that predicted for an entangled phase of living polymers. A qualitative mechanism that combines the effect of {\csd} on average micelle length and Debye parameter with shear-induced configurational changes of rod-like micelles is proposed to rationalize the self-similarity of SIS formation.

\end{abstract}

\section{INTRODUCTION}

Surfactant solutions exhibit complex behavior even at equilibrium.
Above a critical micelle concentration, several tens of surfactant molecules aggregate to form micelles since the formation of micelles is entropically favorable as compared to a system of unassociated surfactant molecules in solution {[}Hamley (2000)]. Micelles could exist in various shapes such as cylindrical, spherical or lamellar depending on the temperature and surfactant concentration in solution {[}Hamley (2000), Tsujii (1998), Tardos (1984), Zana (1987), Russell {\it {et al.}} (1989), and {\Shen}]. The equilibrium phase behavior, however, could be qualitatively influenced by flow. Understanding the relationships between flow, phase behavior, and rheology is central to numerous applications. Examples include (wormlike) micelles-induced turbulent drag reduction, drug delivery, alveolar function, synthesis of nanoporous materials (e.g. sol gel process), enhanced oil recovery and solar energy storage {[}{\Tardos}, {\Zana}, {\Russel}, {\Tsujii}, {\Hamley}, Arora {\it {et al.}} (2002), and {\Shen}].

Rheological behavior of surfactant solutions is very sensitive to the concentrations of the surfactant and counter-ion, temperature, and the presence of added salts. Relatively small changes in these parameters can alter micelle shape/size and, in-turn, the rheology. Since micelles of ionic surfactants are charged, addition of counter-ions decreases the Debye screening length $\kappa^{-1}$, thereby reducing the intermicellar repulsive forces and favoring the formation of larger micelles. While spherical shapes are known to be preferred for small surfactant concentration ($C_{D}$), cylindrical and lamellar shaped structures are generally observed as $C_{D}$ is increased {[}{\Cates}]. At sufficiently large $C_{D}$ values and/or under flow, wormlike micelles that behave similar to flexible polymers can be formed {[}{\Hamley}, {\Cates}]. The contour length of such micelles could be as large as several microns {[}{\Clausen}, Lin {\it {et al.}} (1992, 1994, 2001)]. The length of such micelles depends on the energy required to form the two chain ends that is typically ${\approx}$ 5-25 $kT$ {[}{\Catesb}], which is also referred to as the scission energy $\alpha$. If $\alpha$ ${\gg}$ 25 $kT$, very long micelles are formed that could entangle with one another even at relatively low volume fractions {[}{\Cates}, {\Oelschlaeger}]. Such entangled phases could exhibit gel-like characteristics {[}{\Rehage}, {\Nowak}, {\Shikata}].
Under shear, the scission energy $\alpha$ could be overcome and the end caps could be broken, resulting in micelles with residual charges. Moreover, wormlike micelles in shear flow can undergo stretching-tumbling cycles. In a stretched (extended) state, the probability of inter-micellar interactions is enhanced compared to that in the coiled (equilibrium) state. Hence, under flow, micelles with residual charges can combine to form longer micelles or networks {[}{\Cressely}]. This effect is more pronounced for micelles with length $L$ larger than the persistence length $l_{p}$, which behave like flexible polymers {[}{\Cates}].

The pioneering work by {\Cates} allowed for a quantitative description of the rheological behavior of wormlike micellar solutions. A vast majority of wormlike micelles in dilute solutions exhibit linear viscoelastic behavior such as the one predicted by the linear Maxwell model with a single relaxation time {[}{\Rehage}; {\Kim}; {\Shikata}] or nonlinear shear-thinning behavior similar to a Giesekus fluid {[}Shikata and co-workers (1987a, 1987b, 1988)]. However, under certain salt to surfactant concentration ratios, they exhibit shear thickening behavior, i.e. an increase in the apparent viscosity of the solution with increasing shear rates {[}{\Hub}; Cressely and co-workers (1994, 1997a, 1997b, 1997c, 1998a, 1998b, 2000); {\Wheeler}; {\Nowakb}]. In a series of experiments, Cressely and co-workers (1994, 1997a, 1997b, 1997c, 1998a, 1998b, 2000) studied the effect of counter-ion concentration $C_{s}$, surfactant concentration $C_{D}$, different added salts, and temperature on the rheology of the salt-surfactant system. They reported that the rheological behavior differs dramatically for different types of salts. In the range of surfactant [Cetyl-trimethylammonium bromide (CTAB)] concentrations they studied, shear thickening behavior was observed with addition of counter-ions such as Sodium Salicylate (NaSal) and Sodium Tosylate (NaTOs). However, shear thickening behavior was not observed while adding Sodium Benzoate (NaBz), even though the three salts were chemically similar except for their polar head groups. This behavior was attributed to the differences in the strength of the counter-ion - surfactant bond as well as the placement of the counter-ions on the surfactant {[}{\Hartmannb}]. {\Cates} have noted that scission energy is strongly dependent on the type of counter-ion, which in turn would control the length of the micelles formed and hence the solution rheology. Goyal and co-workers (1989, 1991, 1992) studied the effect of Pottasium Bromide (KBr) as well as NaSal on the zero-shear viscosity (${\eta_{0}}$) of CTAB solutions. Both these salts produced a large change in ${\eta_{0}}$, however, for isoviscous micellar solutions of CTAB/KBr and CTAB/NaSal,  KBr micelles were found to be longer than NaSal micelles. They attributed this behavior to the Sal$^{-}$ ions being absorbed on the surface of micelle as a cosurfactant, unlike Br$^{-}$ ions, which remained in the bulk of the solution.

The rheological behavior for various $C_{D}$ values was investigated by {\Cressely} with the CTAB/NaSal system. They fixed {\csd} = 0.019M and changed {\cds} (0.045M $\leq${\cds} $\leq$ 0.3M), and observed that {\zsv} initially decreased nonlinearly from 0.9 Pa.s ({\cds} = 0.045M) and plateaued at 0.004 Pa.s ({\cds$^{*}$} $\approx$ 0.15M). Further increase in {\cds} resulted in a gradual increase in {\zsv} to 0.05 Pa.s ({\cds} = 0.3M). Thus, for a given {\csd}, an increase in {\cds} by a factor of seven resulted in two orders of magnitude variation in {\zsv}. {\Lin} suggested that the initial decrease in {\zsv} was possibly due to either the formation of micellar branches (which increased the fluidity of the solution), micellar bundles (which moved cooperatively as single objects), and  the saturation of micellar networks depending on the molar ratio of salt to the surfactant (\emph{R} = {\csd}/{\cds} ). They suggested that the subsequent increase in {\zsv} for {\cds}$>$\cds$^{*}$  was due to the formation of many more micellar bundles. {\Azzouzi} analyzed {\zsv} data for fixed {\cds}  = 0.05M and {\csd} varying from 0.01M to 1M. They observed that the rheological behavior of the solution dramatically changed as {\csd} increased in this region, and the shear thickening regime was limited to a narrow range of {\csd} values. Only in the range of 0.01M < {\csd} < 0.018M, both {\Azzouzi} and {\Cressely} reported shear thickening behavior. For higher {\csd} values, {\Azzouzi} reported a domain of heterogenous flows and shear induced phase transitions that exhibited shear thinning behavior. In the present study, CTAB/NaSal solutions have been prepared for various salt concentrations that lie within the shear thickening regime, and hence the salt concentrations are below 0.019M for the case of {\cds} = 0.05M. Further details of the sample preparations are given in Section II.

{\Kim} studied the rheology of CTAB/NaSal solutions with 0.1 < \emph{R} < 10 and concluded that \emph{R} strongly influences the viscoelastic properties of the solution. Using a single relaxation time ${{\tau}_{R}}$ to fit the linear viscoelastic response of solutions to the Maxwell model, they observed that ${{\tau}_{R}}$ decreased with increasing \emph{R} for 1 < \emph{R} < 10. However, the plateau modulus $G{}_{0}$ was independent of \emph{R}. In our study, we focus on shear thickening solutions of CTAB/NaSal in the range 0.18 < \emph{R} < 0.34, and also obtain an estimate of  the gel modulus that is  independent of {\R}. We compare our results to those reported by {\Kim}, see Section III for details.

While the rheology of surfactant-salt systems has been studied extensively, the formation and morphology of Shear Induced Structures (SIS) have also been the subject of several investigations. Specifically, rheo-optical techniques have been used extensively to study the evolution and microstructure of SIS, in shear thickening systems. {\Wunderlich} performed both birefringence and rheological studies on tetradecyl trimethyl ammonium salicylate (TTAS) solutions and inferred that the SIS consist of small clusters of rod-like micelles in the shear-thickening regime. Small angle light scattering (SALS) measurements on the CTAB/NaSal system were reported in the pioneering work by {\Liu}, wherein they observed an increase in the shear stress as the shear thickening transition occurred. Using a CCD video camera to visualize the fluid flow in a Couette cell, they observed finger-like structures that exhibited gel-like characteristics. The gel phase extended from the stationary inner wall to the rotating outer wall of the Couette cell. In the shear thickened state, they also observed fluctuations in the stress that were attributed to the growth and retraction of these shear induced fingers due to an elastic interfacial instability. Their results are consistent with the findings of Bandyopadhyaya and co-workers (2000, 2001) who observed time dependent behavior of shear as well as normal stresses in the SIS regime above a critical shear rate for a cetyltrimethylammonium tosylate (CTAT) solution, and obtained a positive Lyapunov exponent for the temporal response of the stress signals, implying large amplitude fluctuations. Chen and Rothstein (2004) observed velocity fluctuations and a turbid high-shear region around a sphere falling in a micellar solution, and conjectured that these effects were due to the formation of a shear induced gel-like state. Very recently, {\Ouchi} used birefringence and optical visualization (with tracer particles) to study SIS in the start-up Couette flow of CTAB/NaSal solution. They observed that an initially transparent fluid became opaque because of SIS formation.

Using light scattering experiments to study the shear thickening regime, {\Protzl} showed that SIS formation can be divided into three consecutive stages, namely, induction, aggregation, and orientation of micelles. With the same technique, {\Hu} attributed the origin of the shear thickening regime specifically to heterogenous nucleation of SIS because of the presence of impurities and/or surface roughness. Using NMR velocimetry and mechanical measurements within a cone-and-plate rheometer, shear banding, i.e. formation of two layers with high contrast in shear rates, was observed by Britton and Callaghan (1997,1999) in the CPyCl/NaSal system. Moreover, in the shear thickened regime, they observed different shear rate phases separated by a boundary that either exhibited rapid fluctuations or slow migrations in the direction of the velocity gradient. The fluctuation and migration effects were attributed to interfacial instability and stress relaxation respectively. Chen and Rothstein (2004) also reported velocity fluctuations of a sphere falling in a micellar solution, and observed significant rotation and migration of the sphere towards the confining walls, and attributed these effects to an instability in the SIS phase.

Several authors have attempted mathematical modeling of the rheological behavior of surfactant-salt systems to address the factors that govern the formation and dynamics of SIS (e.g. Cates and co-workers (1987a, 1987b, 1988, 1990, 1996, 2001), {\Shankar}, {\Shankarb}, Bautista (1999), Maci$\acute{}$as \emph{et al.} (2003), McLeish (1986, 1987), {\MacKintosh}, {\Appell}). Special attention has been paid to the formation of shear bands by Olmstead  and co-workers (2003, 2004, 2006).  The dynamics of the interface between the gel-like and solution phases have been modeled by {\Shankar} and {\Kumaran}. Kumaran and co-workers {[}{\Kumaran}, {\Kumaranb}, {\Kumaranc}] performed linear and nonlinear stability analysis of a Couette flow of a Newtonian fluid with viscosity $\eta$, past a flexible gel-like medium with shear modulus {\G} in the limit of negligible inertial effects. The analysis revealed that an instability can occur at the fluid-gel interface above a critical value of the dimensionless stress parameter  $\Gamma = {\eta \dot{\gamma}}/{\emph{G}}$. Second, the critical value of $\Gamma$ above which instability occurred depended on the ratio of the thickness of the gel and fluid phases, ratio of the gel and fluid viscosities, as well as interfacial tension. {\Kumar} applied this model to interpret experimental data for surfactant solutions in the SIS regime, where the structures were predicted to exhibit gel like properties {[}{\Hu}].  They observed an instability in their fluid-gel model and reported a critical dimensionless stress $\Gamma_{C}$ for the onset of instability.  Comparison of theoretical and experimental values suggested that $\Gamma_{C} \approx$ 0.3. In the present study, both the fluid-gel model and $\Gamma_{C}$ values have been used to estimate the elastic properties of the gel-phase: see section III.H.

Despite the numerous studies described above, a fundamental first principles based understanding of the relationship among flow, electrostatics and microstructure is only slowly emerging for SIS forming systems. Specifically, electrostatic interactions are greatly dependent on the salt concentration. First principles modeling efforts should identify and faithfully incorporate salt mediated interactions to have truly predictive capabilities. Evidently, precise knowledge of the effect of salt concentration on the critical shear rates and inception times (strains) is a pre-requisite for gaining physical insights and validation of models. In this work, we have focused on performing a series of experiments for a well-characterized system (CTAB/NaSal) [Cressely and co-workers (1994, 1997a, 1997b, 1997c, 1998a, 1998b, 2000), {\Liu}, {\Azzouzi}, {\Kim}] in the shear thickening regime with the objective of identifying the effect of {\csd} on the onset conditions for SIS formation. Based on the experimental data, a scaling relationship has been proposed for {\csr} vs. {\csd}. Moreover, it is shown that for a given {\csd}, all {$\eta$} vs. {\sr} data for shear thickening solutions can be reduced to a single master curve, suggesting a self similar dynamical behavior. In addition to the above findings, we have shown that shear thickening transitions in the above system are indistinguishable in terms of {$\eta$} vs. {\sr} data obtained from two geometries, namely the cone-and-plate and the Couette flows. We also apply interfacial instability theories for model gel/fluid systems to obtain estimates of the elastic modulus of the gel phase and compare them to independent measurements.

The rest of this manuscript is organized as follows : materials and
methods are discussed in II, results are presented and discussed in
III and conclusions are given in IV.

\section*{II MATERIALS AND METHODS}

As stated above, our objective is to perform rheological characterization that will allow us to quantify the effect of {\csd} on the onset conditions of SIS formation. Hence, we have used CTAB/NaSal as the surfactant - salt system in our studies since several rheological and optical measurements of this system are available in the literature. The samples of CTAB and NaSal were obtained from Fluka and Sigma Aldrich respectively. All the solutions were prepared using deionized water with resistance greater than 17.9 $M\Omega$ and then equilibrated for three days before being used for analysis. After master batches of CTAB and NaSal were prepared, calculated volumes were stirred and diluted to achieve the desired concentrations. Three sets of samples with surfactant concentrations {\cds} = 0.05M, 0.07M, and 0.09M were prepared with molar ratios {\R} = $C_{s}$/$C{}_{D}$ varying from 0.18 to 0.34 in each set. These concentrations were chosen so that  the shear thickening transitions   obtained by {\Cressely} could be reproduced. The samples were stored in the laboratory at room temperature without direct exposure to sunlight.

Rheological experiments were performed using TA Instruments AR 2000 constant stress and strain rheometer, primarily using a cone and plate geometry with a diameter of 60 mm (cone angle 0° 59' 49\char`\"{}, truncation 27 $\mu$m). A set of experiments was also performed using the Couette geometry to ensure the robustness of the data. The torque range of the rheometer was 0.1 $\mu$Nm - 200 $\mu$Nm, and the minimum stress measurable was 0.0008 Pa. Temperature was controlled at 24$^\circ$C by  a Peltier plate. A solvent trap was used to minimize evaporation. At the end of each experiment, the Peltier plate was flushed with water to remove the sample and then dried and cleaned using Isopropyl alcohol before adding a new sample. The Couette experiments were performed with standard-size double concentric cylinders with an outer aluminum rotor (OD 21.96 mm, ID 20.38 mm) immersed at a height of 500 ${\mu}$m inside the inner steel stator (ID 20 mm). The solvent trap was also used for this instrument to minimize evaporation.

In order to determine an optimum sampling time per data point, steady shear experiments were performed with different sampling times. It was found that 600 seconds was sufficient to obtain reproducible data for $\eta, {\tau}$ and $\psi_{1}$, as discussed in section III. Three trials were performed for each set of ($C_{D}, C_{s}$) to ensure the reproducibility of results and to estimate the experimental uncertainties.

Samples stored for over 4 months showed aging effects with change in {\zsv}  and/or exhibited shear thinning behavior. Similar aging behavior was also observed by {\Mohan}, {\Myska}  and {\Jana}. Shear thickening solutions with high  molar ratios exhibited shear thinning on aging, while those with low molar ratios continued to shear thicken, however accompanied by a decrease in {\zsv} and/or an increase in the critical shear rate {\sr$_{c}$}. The former effect was attributed to the formation of networks upon aging {[}{\Jana}], and the latter could be due to the degradation of the micelles at equilibrium below a certain \emph{R}. Some of our samples also showed photosensitive effects by turning brown upon aging. However, all the data reported here are for samples less than 60 days old for which aging effects are absent.

\section*{III RESULTS AND DISCUSSION}

\subsection*{A. Specific viscosity and micelle structure}

The effect of salt concentration {\csd} on the specific viscosity $\eta_{sp}$ of the solutions near equilibrium was analyzed. Here, $\eta_{sp}$ $\equiv$ ({\zsv} - $\eta_{s}$)/$\eta_{s}$, where {\zsv} and $\eta_{s}$ denote the zero-shear viscosity of the solution and the solvent respectively. The variation of $\eta_{sp}$ as function of {\csd} is given in Fig. \ref{figure1}. The values of the surfactant concentrations  {\cds} considered here are 0.05M, 0.07M and 0.09M, with molar ratio of salt to surfactant concentrations \emph{R} $\equiv$ $C{}_{s}/C_{D}$ varying from 0.18 to 0.34 in increments of 0.02M in {\csd} (denoted by numbers from 1 to 9). The scaling exponents for the three sets of data for $\eta_{sp}$ versus {\csd} are within experimental uncertainty, and range from 11.6 to 12.4. We observe over three orders of magnitude increase in $\eta_{sp}$ corresponding to a factor of two enhancement in {\csd}, signifying that the microstructure is extremely sensitive to variations in {\csd}.

The inset a in Fig. \ref{figure1} shows the log-log plot of {\zsv} with respect to {\csd}. For {\cds} = 0.05M, {\zsv} initially increases almost linearly with increasing {\csd} and thereafter shows a power-law behavior. The {\zsv} values reported here are in agreement with those reported by {\Azzouzi} and {\Cressely}. Similar change in the scaling behavior can also be observed for solutions with {\cds} = 0.07M, while all the solutions with {\cds} = 0.09M exhibit a power-law behavior. The trends observed for {\cds} = 0.05M and 0.07M indicate a possible change in the microstructure as {\csd} increases. For any fixed {\cds}, the changes in {\zsv} with respect to {\csd} are representative of the micellar contribution to the viscosity of the solution. It is well known that globular micelles form at low {\cds} and/or with no salt {[}{\Hamley}]. Increasing the salt concentration reduces the electrostatic Debye screening length around the surfactant, which encourages the formation of longer micelles at equilibrium. This in turn contributes to the changes in $\eta_{0}$. {\Fujio} found that spherical micelles associated to form into rod-like micelles when {\csd} exceeded a threshold concentration.

{\Imaeb} studied the sphere-to-rod transition of CTAB/NaBr micelles, and showed that the logarithm of micelle molecular weight (MW) $\sim$  log($C_{0}$ + {\csd}){$^{n}$}, where $C_{0}$ is the critical micelle concentration, and the exponent \emph{n} < 1 for spherical micelles, whereas \emph{n} > 1 for rod-like micelles. Similar observations have been made for other micellar systems as well [{\Fujio}, {\Imae}]. To verify the structure transition in our study, we need to estimate both MW and $C_{0}$ of the micelles. To calculate MW, we referred to the lengths of CTAB/NaSal micelles obtained by {\Garg} who used Dynamic Light Scattering (DLS) measurements to study the structure of the micelles with {\cds} = 0.05M and 0.015M < {\csd} < 0.05M. Their DLS data was analyzed in terms of a prolate ellipsoidal structure with the semi-minor axis of the ellipsoid fixed at 4.3nm. The semi-major axis was used to obtain the average length of the micelles ($\bar{L}$) for the different {\csd} values, see Table 2, {\Garg}, from which $\bar{L}$ = 20  $\pm$ 2 nm for {\cds} = 0.05M and {\csd} = 0.015M, and increases to 93 $\pm$ 9 nm for {\cds} = 0.05M and {\csd} = 0.05M. {\Imae} have reported the average micellar molecular weight for cylindrical micelles = $\bar{L}$M{$_{L}$}, where M{$_{L}$} is the molecular weight per unit length $\approx$ 8900 nm$^{-1}$ for CTAB/NaSal system [{\Shikatad}]. 

The critical micelle concentration $C_{0}$ of CTAB in pure water is 9 $\times$ 10$^{-4}$ M [{\Rothstein}]. Addition of salt is known to decrease $C_{0}$ of the solution [{\Hamley}]. Hence, when NaSal is added to the solution, $C_{0}$ < 9 $\times$ 10$^{-4}$ M. In our study, both {\csd} and {\cds} $\gg$ 9 $\times$ 10$^{-4}$ M, hence $C_{0}$ + {\csd} $\approx$ {\csd}. Based on the estimates discussed above, a plot of log MW vs. log {\csd} is shown in the inset 1b. A significant change in slope can be seen around {\csd} $\approx$ 0.022M (log {\csd} $\approx$ -1.65) which indicates predominantly spherical micelles for lower {\csd} values, and more rod-like micelles for higher {\csd} values. Using Cryo-TEM technique on CTAB/methylsalicylic acid, {\Linb} have shown that both spherical and wormlike (or rod-like) micelles coexist for a range of salt-surfactant concentrations.  Thus a large size distribution of micelles including spherical and polydisperse rod-like micelles should exist in this regime. Since globular micelles form at low salt and surfactant concentrations {[}{\Hamley}],  we hypothesize that at low {\R}, there are more spherical micelles than rod-like micelles, and as {\R} is increased beyond 0.26 ({\csd} = 0.013M),  the number of polydispersed rods exceed the number of spheres. For a given {\R}, we see that an increase in {\cds} from 0.05M to 0.09M leads to an increase in {\zsv}, which suggests that longer rod-like micelles could form with the addition of surfactant and salt. Whereas, for a given {\csd}, a change only in {\cds} does not necessarily promote the formation of longer micelles. This can be seen from the inset b in Fig. \ref{figure1}, where for a given {\csd}, {\zsv} decreases as {\cds} increases from 0.05M to 0.09M. This trend was also observed by {\Cressely}, who attributed the decrease in {\zsv} to the relative decrease in the amount of salt per unit mass of the surfactant, which hindered the formation of longer rod-like micelles.

From {\Garg}, the length for micelles with {\cds} = 0.05M, and {\csd} = 0.015M $\approx$ 20  $\pm$ 2 nm. This specific datum falls within the parameter range of the present study for which {\cds} = 0.05M and 0.009M < {\csd} < 0.019M. Using the theory for uncharged rigid rods derived by {\Doi}, we found that the average distance between the rods in the suspension (${\nu}^{-1/3}$) was larger than the rod length $\bar{L}$, i.e. ${\nu}\bar{L}^3 \leq 1$, implying a dilute regime.

\subsection*{B. Shear rheology and critical shear rate}

In Fig. {\ref{figure2}}, we show the results of a steady shear experiment using the solutions with {\cds} = 0.05M and 0.18 $\leq$ {\R} $\leq$ 0.34. For small values of {\it{R}} and at shear rates {\sr} < 500 s$^{-1}$, the homogeneous solution has viscosity ${\eta}_{app}$ comparable to that of the solvent  (${\approx}$ 1 mPa.s). However, as the shear rate is increased beyond a threshold value {\csr}, e.g. approximately 1580 s$^{-1}$ for {\R}  = 0.18, SIS formation is observed. SIS manifest as a gel-like micellar phase dispersed within a relatively low viscosity solution. As \emph{R}   is increased, $\eta_{0}$ and the maximum in the viscosity ${\eta}_{app}$ increase, while {\csr} decreases. From Fig. {\ref{figure2}}, it is evident that as {\csd} (or \emph{R}) increases, the relative magnitude of shear thickening gradually decreases such that when the molar ratio {\R} = 0.34, the solution shear thins. As {\R} is increased, electrostatic interactions are more effectively screened. This favors the formation of longer micelles which behave under flow as flexible macromolecules. This is consistent with the shear thinning behavior observed for large {\csd} values. The findings reported in Fig. {\ref{figure2}} are also consistent with the results of {\Hartmannb}.  The dependence of the critical shear rate for the onset of shear thickening {\csr} with respect to {\csd} is given in Fig. {\ref{figure3}}, and we see that {\csr} {$\sim$} {\csd}$^{-6}$. This is in qualitative agreement with the results of {\Hartmannb} whose data showed {\csr} {$\sim$} {\csd}$^{-7}$. Note that while data reported in the present study are averaged over three runs, {\Hartmannb} have not reported standard deviations for their data. The determination of critical shear rate has to be done judiciously considering the experimental uncertainty incurred in estimating the apparent viscosity (as an average of a fluctuating signal over a sufficiently long period of time: see section III.C. below) and by visual inspection of the data to locate an abrupt increase in $\eta_{app}$. This procedure when applied consistently, translates to identifying {\csr} as the shear rate at which ($\eta_{app}$ - {\zsv})$/$ {\zsv} $\geq$ 0.1. {\Hartmannb} do not report the criterion used to determine {\csr} in their work. Since the shear-thickening transition is abrupt and very sensitive to {\sr} in the vicinity of {\csr}, such uncertainties can lead to differences in best fits to the data. 

\subsection*{C. Torque fluctuations}

In the SIS regime, the interface between the gel phase and the surrounding fluid phase oscillates with a frequency that depends on the salt concentration {[}{\Callaghan}]. The interface may develop fingers which grow and retract {[}{\Liu}], and migrate slowly in the direction of the velocity gradient {[}Britton and Callaghan (1997, 1999)]. Such unstable interfacial dynamics cause torque fluctuations, and a small variation in {\csd}, {\cds} and/or the presence of small impurities can substantially alter the dynamics as shown by Britton and Callaghan (1997, 1999). These authors conjectured that the interfacial boundary moved by ejecting or acquiring fluid of the two phases resulting in normal stress discontinuity and slow migration of the interface. In Fig. {\ref{figure4}}, the oscillatory, but statistically stationary torque signals obtained from the rheometer for a solution with {\cds} = 0.05M and {\R} = 0.2 are illustrated. Here, the torque signals have been recorded for 1800 seconds in the SIS regime by performing a Peak Hold in which the shear rate was held at a constant value over a certain time period. Since the torque signals are oscillatory, the viscosity has to be interpreted as an apparent value ${\eta}_{app}$, i.e., as the ratio of the time-averaged shear stress to the shear rate. The inset in Fig. {\ref{figure4}} displays ${\eta}_{app}$ obtained at the respective {\sr} values upon time averaging the torque signals over 600 s (solid line) and over 1800 s (symbols). Since the difference between these values is very small, 600 seconds was chosen as sufficient run time for steady shear experiments. Fig. {\ref{figure4}} also shows that in the SIS regime, as {\sr} increases, the torque fluctuations increase. In Fig. {\ref{figure5}}, the root-mean-square values of these fluctuations are shown for two solutions, {\cds} = 0.05M ({\R} = 0.2) and {\cds} = 0.07M ({\R} = 0.22). In both  cases, the magnitudes of fluctuations in the shear-thickened regime reach values that are at least an order of magnitude larger than those in the Newtonian regime. This result suggests the chaotic nature of SIS. This dynamic behavior is consistent with the results of {\Liu}, {\Nowak}, and Bandyopadhyay and co-workers (2000, 2001).

\subsection*{D. Critical strain}

In the above analysis, a step increase in {\sr} is accomplished without any stoppage in flow, i.e., the strain is continuously applied on  the solution. However, starting at equilibrium, if a {\sr} $\ge$ {\csr} is applied on the solution, the stress (or torque) increases after some time period {[}{\Liu}]. The time needed for the stress to rise, decreases as {\sr} is increased in the shear thickening regime. This is illustrated in Fig. {\ref{figure6}}, where for a solution with {\cds} = 0.05M and {\R} = 0.28, we applied shear rates of 100, 150, 200 and 300 s$^{-1}$, denoted by the filled symbols in the figure. The inset shows $\eta_{app}$ vs. {\sr}. Based on the time $t^{*}$ required for the increase in stress, we obtain the critical strain ($\gamma_{C}^{*}$ = {\sr}$t^{*}$). Irrespective of the magnitude of {\sr} applied, $\gamma_{C}^{*}$ was found to be almost constant $\approx$ 8500. Hence the shear thickening transition occurs when a critical strain is imposed on the micellar solutions at equilibrium, as long as the {\sr} $\geq$ {\csr}. We found that the value of the critical strain $\gamma_{C}^{*}$ was dependent on {\csd}. For example, in the case of solutions with {\cds} = 0.05M, $\gamma_{C}$ $\approx$ 8500 for {\csd} = 0.014M, whereas $\gamma_{C}$ $\approx$ 5750 for {\csd} = 0.015M. This is because of the more effective screening of electrostatic interactions that leads to longer micelle length under flow for larger {\csd} values. Since longer micelles can entangle more easily, smaller strains are required to form SIS.

\subsection*{E. Stress relaxation and normal stress}

When the strain applied on the shear induced solution was stepped down to zero, the stress decreased rapidly and abrupt at first, as shown in Fig. {\ref{figure7}}. This rapid decay is followed by a slower stress relaxation process. This bimodal relaxation process agrees well with the relaxation behavior of entangled solutions of threadlike micelles [see {\Amin}, {\Shukla}, {\Cates} and references therein], and corroborates the understanding that SIS are entangled phases. Thus, the stress relaxation mechanism is governed by a combination of local cooperative motions of the gel-like network of SIS, and relaxation of the transient network structure. For the given solution (Fig. {\ref{figure7}}), it is observed that the shear stress decays by over four orders of magnitude, and for long times, the residual stress B ($\approx$ 0.0027 Pa) is statistically stationary for the duration of the experiment (300 s in this case). Experiments conducted for over 10 hours also yielded this value of B. When the stress decayed to about 10\% of its initial value, the data was fit to a single exponential model for the stress, $\tau = A exp(-t/\lambda_{\tau}) + B$, where $\lambda_{\tau}$ is the longest stress relaxation time, and A is a constant. The fit is shown by the solid line in Fig. {\ref{figure7}}. $\lambda_{\tau}$ depends on the counter-ion concentration. For example, $\lambda_{\tau}$  $\approx$ 2.1 s and 2.6 s for {\R} = 0.28 and 0.30 respectively, for a given applied strain $\gamma$ = 9556. As the counter-ion concentration increases, the enhanced electrostatic screening could promote the formation of more entangled network which would relax slower.

The SIS regime is known to exhibit viscoelastic behavior {[}{\Nowak}]. SIS are attributed to flow-induced deformation of the gel phase that manifests as elastic normal stresses. In the shear induced regime, {\Fischer}  reported oscillations of both the shear stress ($\tau$) and first normal stress difference $N_{1}$ to be in-phase, signifying that these oscillations  were caused by the same mechanism. Second, the transient but pronounced behavior of normal stresses indicated that SIS regime was more viscous and elastic than the Newtonian phase {[}{\Fischer}]. In Fig. {\ref{figure8}}, we show the apparent normal stress coefficient {\nsc} as a function of {\sr} for a typical shear thickening solution with {\cds} = 0.05M and {\R} = 0.24. Note that {\nsc} can be measured only in the shear-thickened regime, which proves that the SIS are elastic in nature. This result agrees with the findings of {\Wunderlich}, Hu and Matthys (1995, 1996), and {\Nowakb}, who have also reported that the normal stress effects are observable only in the SIS regime. However, as seen from Fig. {\ref{figure8}}, the shear rates for the onset of shear thickening in viscosity ({\csr}) and  for the development of observable  normal stresses do not coincide. These observations indicate that although the entangled phase contributes to the increase in {\appvisc}, the fluid becomes more gel-like with time perhaps due to consolidation of the network structure. The normal stresses grow gradually, plateau and finally decrease at higher shear rates. The plateau indicates that the gel structure is robust within a certain shear rate range. At higher shear rates, the partial destruction of aggregates decreases the normal stresses, for instance, for {\R} = 0.2, {\nsc} decreases beyond 3000 s$^{-1}$ {[}{\Nowak}]. Above {\sr} > 5000 s$^{-1}$, we  observed foaming, and centrifugal forces throw the fluid out of the rheometer. Hence our analysis is restricted to shear rates $\le$ 5000 s$^{-1}$.

\subsection*{F. Self-similar transition to SIS regime}

We define an apparent relaxation time as $\lambda (\dot{\gamma}) \equiv \left.{\frac{\psi_{1}}{2\eta_{app}}}\right|_{\dot{\gamma}}$ and  measured the effect of {\csd} on $\lambda$  for all cases in which shear thickening transition is observed. In Fig. {\ref{figure9}}, we plot the largest relaxation time $\lambda$ ($\equiv \left.{\lambda}\right|_{\psi_{1,max}}$, where ${\psi_{1,max}}$ is the maximum normal stress coefficient)
as a function of {\csd}, and observe that {$\lambda$} $\sim$ {\csd}$^{6}$.  The product of {$\lambda$} and the {\csr} ($\sim$ {\csd}$^{-6}$, see Fig. {\ref{figure3}}) is practically independent of the salt concentration, since the scaling exponents of {\csr} and $\lambda$ are -6 and 6 respectively. We define a Weissenberg number \emph{We} $\equiv \left.{\lambda}\right|_{\psi_{1,max}} \left.{\dot{\gamma}}\right|_{\psi_{1,max}}$, where $\left.{\dot{\gamma}}\right|_{\psi_{1,max}}$ is the shear rate at which ${\psi_{1,max}}$ is obtained. As shown in Fig. {\ref{figure9}}, \emph{We} is only weakly dependent on the salt concentration. This suggests a self-similar dynamical transition to the SIS regime. Taking advantage of this observation, a master plot of viscosity vs. shear rate may be constructed by suitably shifting the coordinates for all shear-thickening cases.  We use the following definitions for a reduced viscosity ($\eta_{R}$), reduced shear rate ($\dot{\gamma}_{R}$), and the normalized reduced viscosity ({${\eta}_{NR}$}) to normalize the viscosity data for a solution (\emph{S}) with respect to that of a base solution (\emph{B}).

\begin{equation}
\eta_{R}\equiv\frac{\eta}{A_{c}} ,\label{eq:1}\end{equation}

\begin{equation}
\dot{\gamma_{R}}\equiv\dot{\gamma}A_{c} ,\label{eq:2}\end{equation}

where

\begin{equation}
A_{c}=\frac{\eta_{0,S}}{\eta_{0,B}} ,\label{eq:3}\end{equation}

and

\begin{equation}
\eta_{NR}=\eta\frac{\eta_{0,B}}{\eta_{0,S}}\frac{\eta_{max,S}}{\eta_{max,B}} .\label{eq:4}\end{equation}

In the above equations, $\eta{}_{max}$ denotes the maximum apparent viscosity in the SIS regime. The master curve for  {\cds} = 0.05M is given in Fig. {\ref{figure10}}, where $\eta_{NR}$ is plotted versus the reduced shear rate ($\dot{\gamma_{R}}$).

Note that the power-law scaling behavior where {\csr} $\sim$ {\csd}$^{-6}$ and $\lambda \sim $ {\csd}$^{6}$ is observed only for salt concentrations greater than a critical value (which slightly exceeds the minimum salt concentration required to induce shear thickening) and, less than the {\csd} value for which elongated micelles are formed at equilibrium causing shear thinning behavior. Only data for this concentration range are used in the renormalization procedure. This corresponds to R values in the range of 0.24 - 0.32 in this case of {\cds} = 0.05M. Here, we chose {\R} = 0.24 as the base case (\emph{B}). The selection of the base case does not affect the master plot.

In order to validate whether the above conclusions are robust with respect to surfactant concentration, experiments with {\cds} = 0.07M and 0.09M, and with 0.18 < {\R} < 0.34  were performed. In Figs. {\ref{figure11}} and {\ref{figure12}}, we plot $\eta_{app}$ vs. {\sr} for {\cds} = 0.07M and 0.09M respectively. Solutions with {\cds} = 0.07M and 0.09M also show shear thickening behavior, though the range of {\R} values showing this behavior decreases as {\cds} increases.  For {\cds} = 0.07M,  solutions with 0.18 $\le$ {\R} $\le$ 0.24 showed shear thickening behavior, while for {\cds} = 0.09M, the shear thickening range was 0.18 $\le$ {\R} $\le$ 0.22.  Fig. {\ref{figure13}} shows the best power-law fits for {\csr} and $\lambda$ with respect to {\csd}, for {\cds} = 0.07M and 0.09M respectively. The  \emph{We}  for the respective salt concentrations is also given, and it is evident that \emph{We} is weakly dependent on {\csd}. For these shear thickening solutions, {\csr} $\sim$ {\csd}$^{-6}$ and $\lambda$ $\sim$ {\csd}$^{6}$  respectively, similar to {\cds} = 0.05M. The specific exponents for {\csr} and $\lambda$ are -5.8 and 6.2 for {\cds} = 0.05M, -6 and 5.9 for {\cds} = 0.07M, and -5.8 and 6.1 for {\cds} = 0.09M. Thus the scaling behavior of {\csr}, $\lambda$, as well as the dependence of \emph{We} on the counter ion concentration is robust and is almost irrespective of {\cds}.

Figs. {\ref{figure14}} and {\ref{figure15}} show the master plots for {\cds} = 0.07M and 0.09M respectively, obtained using Eqs. \ref{eq:1} - \ref{eq:4}. For {\cds} = 0.07M, {\R} = 0.22 was chosen as the base case, while for {\cds} = 0.09M, {\R} = 0.18 was selected. Again, the selection of the base case did not affect the master curve. These master curves, however, show some scatter but no systematic trends are observed as a function of {\R}. As discussed earlier, micelles with {\cds} = 0.05M and 0.18 < {\R} < 0.34 exhibited the sphere-rod transition. For a given {\R}, as we increase {\cds}, more micelles could be formed that in turn suggests an increase in intermicellar interaction. Hence, for a given {\R}, {\zsv} increases with {\cds}. Addition of even small amounts of salt screens the electrostatic interactions, and under flow, promotes the formation of longer micelles that tends to shear thin. Hence the number of solutions that exhibit shear thickening transition decreases as {\cds} increases (see Figs. {\ref{figure2}}, {\ref{figure11}}, and {\ref{figure12}}).
Similarly, the extent of shear thickening is reduced as {\csd} increases. However, the master curve suggests that for a given {\cds}, the shear thickening transitions are self-similar.

The mechanism underlying the self-similar behavior must include the salt-mediated screening of the electrostatic interactions among the micelles. Specifically, the Debye layer thickness 1/$\kappa$ vs. {\csd} relationship can be estimated by using $\kappa = (2000 F^{2}/{\epsilon}_{0}{\epsilon}_{r}RT)^{1/2}\sqrt{C_{s}}$, where $F$ is Faraday's constant in coulomb/mol, \emph{R} is the universal gas constant in J K$^{-1}$ mol$^{-1}$, \emph{T} is the absolute temperature in K, ${\epsilon}_{0}$ is the permittivity of vacuum and ${\epsilon}_{r}$ is the relative permittivity of the solvent [Hunter (2001)]. Here, we treat the micelle surface to be continuous with an effective charge density. Fig. {\ref{figure16}} shows 1/$\kappa$ vs. {\csd} for the CTAB/NaSal solutions ({\cds} = 0.05M). In the same figure, the average micelle length $\bar{L}$ is also plotted vs. {\csd} using data reported by {\Garg}. It is seen that $\bar{L} \sim$ {\csd}$^{0.6}$. Hence, as {\csd} is increased, one forms longer micelles with smaller electrical double layers around them. Under flow, the micelles can be stretched, and at relatively high shear rates, the tumbling frequency increases, and hence longer chains tumbling rapidly can form entangled structures. In other words, high shear rates and large {\csd} promote inter-micelle association and hence network formation. Further, in a broad sense, if one views network formation as a chemical reaction, one can interpret the flow-induced conformation change (i.e. stretching/orientation of micelles) as a process that diminishes the associated energy barrier. Hence, the strong dependence of {\csr} on {\csd} can be attributed to the increase in $\kappa$ and $\bar{L}$ with respect to {\csd}. Further, the fact that for a given {\csd}, the strain required to accomplish SIS formation is practically constant implies that for {\sr}  > {\csr}, the inception time decreases as 1/{\sr}. This is qualitatively consistent with the mechanism described above. Specifically, as {\sr} is increased, the chains will unravel faster and their end-to-end tumbling frequency will be increased [Teixeira \emph{et al.} (2005)], which is favorable for network formation. Hence, a mathematical description of SIS formation should combine the knowledge of the physical chemistry of salt-mediated interactions with an analysis of flow-induced configurational changes.

\subsection*{G. Effect of inertia and flow geometry on SIS formation}

To investigate whether the inception times and final state of the SIS could be geometry dependent, we performed steady state experiments  in a Couette cell, details of which are given in section II. In Fig. {\ref{figure17}}, we show the comparison between the Taylor-Couette and cone-and-plate data for two solutions. Both  results are statistically indistinguishable. In the same figure, we report the  Reynolds number \emph{Re}  = $\frac{a^{2}\dot{\gamma}\rho}{\eta}$ vs. {\sr}, where \emph{a} is the characteristic length scale of the flow cell. In the SIS regime, \emph{Re} > 10, which indicates that the inertial effects are significant. However, in the experiments performed by {\Liu} and {\Cressely}, \emph{Re} values were between 0.1 to 120. \emph{Re} values in the present study are similar to those observed in the experiments done by {\Cressely}. This indicates that the SIS formation is not qualitatively influenced by inertial effects.

\subsection*{H. Elastic modulus and network size}

The discontinuity in the elastic normal stress between the gel-like and solution phases in the SIS regime could cause the interface between the two phases to be unstable as in the case of an interface between a deformable gel and a viscous fluid. {\Kumar} have stated that the SIS have  mechanical properties of a gel because these structures are observed to rotate like a solid body and fracture above a critical stress. This has motivated Kumar and coworkers {[}{\Kumar}, {\Kumarb}] as well as {\Kumaran} to study instabilities in model systems such as shear flow of a viscous/viscoelastic liquid over a linear elastic solid. Typically, experiments and stability analysis show that at relatively small Reynolds numbers, interfacial instability occurs when the ratio of the shear modulus of the gel phase {\G} to the characteristic shear stress in the fluid reaches an O(1) value, see Table 1, {\Kumar}. The value of the critical dimensionless stress ${\Gamma_{C}}$ = $\eta${\sr}$/{\G}$, obtained both from experiments and theory was $\approx$ 0.3. Using ${\Gamma_{C}}$ = 0.3 and $\eta$ values at {\sr} = {\sr}$_{max}$, where {\sr}$_{max}$ is the shear rate at which the maximum in ${\eta_{app}}$ is obtained from the present experiments, we have estimated the value of {\G} for the three different surfactant concentrations considered. For {\cds} = 0.05M and  0.22 < R < 0.32, $\eta$ was found to vary $\approx$ {\csd}$^6$. We found that {\G} was  practically constant for different salt concentrations (fixed {\cds}), though it increases with increasing {\cds}. Specifically, {\G}$_{0.05M}$ $\approx$ 5  $\pm$ 0.4 Pa, {\G}$_{0.07M}$ ${\approx}$ 20  $\pm$ 3 Pa, and {\G}$_{0.09M}$ ${\approx}$ 27  $\pm$ 4 Pa. The increase in {\G} with {\cds} can be explained based on the fact that the presence of more surfactants encourages the formation of more networks in the gel phase that in-turn contributes to the increase in the modulus. Independent verification of this gel modulus cannot be performed because the gel formation is reversible i.e. it dissociates upon flow cessation; see Fig. {\ref{figure7}}. However, elastic modulus of CTAB/NaSal gel phases have been measured for R > 1 where micelle networks form even at equilibrium. For instance, {\Kim} used the CTAB/NaSal system with {\R} $\ge$ 1, for which {\zsv} is at least three orders of magnitude greater than those in  our study. They reported a storage plateau modulus {\G$_{0}$} $\approx$ 10 Pa ({\cds} = 0.05M), a value also reported by {\Rothstein} for the same system. Although the equilibrium micellar structures obtained at high {\R} and SIS obtained under lower {\R} values are not readily comparable, the elastic modulus values reported  differ only by a factor of 2 from {\G} $\approx$ 5 estimated by employing linear stability analysis results for a model gel/fluid system {[}{\Kumar}, {\Kumarb}]. Further, we note that the value of 0.3 for ${\Gamma_{C}}$ was obtained from a linear stability analysis for a low Reynolds number flow past polymer gels. Since \emph{Re} in the present system is higher, and  since the model system suggested by {\Kumar}  is not exactly replicated here, the value of ${\Gamma_{C}}$ could differ from 0.3.

If the SIS are regarded as a network of micelles, an estimate of the mesh size ($\xi$) of this network, is given by $\xi\approx(kT/${\G})$^{(1/3)}$ {[}{\Cates}]. This gives $\xi$ values of 100 nm, 60 nm, and 53 nm for {\cds} = 0.05M, 0.07M, and 0.09M respectively. {\Cates} showed that $\xi$ $\sim$ $C_{D}^{-0.77}$ in the semidilute regime which implies that {\G} $\sim$ {\cds}$^{2.3}$. Using the values of {\G} obtained above, we find {\G} $\sim$ {\cds}$^{2.9}$. However, as discussed by {\Cates}, polyelectrolyte effects become important for ionic surfactants at low added salt concentrations. This can lead to a deviation from the theoretical value of 2.3.

\section*{IV CONCLUSIONS}

The influence of  counter-ion concentration {\csd} on the specific viscosity $\eta_{sp}$ and the rheology of dilute CTAB/NaSal system has been investigated to explore whether SIS formation in such systems occur via a self-similar dynamical route. The surfactant concentrations {\cds} considered were 0.05M, 0.07M and 0.09M. The molar ratio {\R} = {\csd} / {\cds} was increased from 0.18 to 0.34 for each {\cds}. For a given {\cds}, it was found that $\eta_{sp}$ was very sensitive to small changes in {\csd}. Using data from the literature [{\Garg}, {\Imaeb}, {\Shikatad}],  the logarithm of the molecular weight (MW) vs. log {\csd} indicated different slopes that suggested the micelles in the concentration regime studied undergo a sphere-rod transition.

Under shear, the solutions exhibited Newtonian behavior at low shear rates {\sr}. When the {\sr} exceeded the critical shear rate {\csr}, a structure transition occurred that caused gel-like structures to form within the Newtonian solvent which in turn increased the viscosity of the solution (shear thickening). As {\csd} increased, {\zsv} also increased, but {\csr} was found to decrease. We found that {\csr} $\sim$ {\csd}$^{-6}$ for all {\cds} values. At high {\csd} values, the solutions exhibited shear thinning behavior. This trend was consistent for solutions with {\cds} = 0.05M, 0.07M, and 0.09M, and for experiments performed in different flow cells, namely the cone-and-plate and Couette rheometers.

The torque fluctuations in the shear thickening regime were found to be at least an order of magnitude greater than the fluctuations in the Newtonian region.  Hence the viscosity values calculated were apparent values ${\eta}_{app}$. When a given {\sr} > {\csr} was applied on the solution originally at rest, the torque was found to increase after a certain time period $t^{*}$. A product of {\sr}$t^{*}$ yielded the critical strain $\gamma_{C}$, which was found to be independent of the magnitude of {\sr} applied on the solution for given {\csd}. However, $\gamma_{C}$ required for SIS formation decreased with increasing {\csd}. When the strain applied on the solution was stepped down to zero, the stress relaxation results suggested that the dissociation time of gel aggregates was also dependent on the amount of counter-ions present in solution. Since SIS formation is reversible, direct measurements of the elastic modulus of gel phase is not feasible. In this work, we have employed results of a hydrodynamic instability theory based on a simplified ``gel-fluid'' model for the heterogenous shear thickened phase and estimated the elastic modulus.

The solutions exhibited both viscous and elastic characteristics only in the shear thickened regime. The ratio of elastic to viscous effects was used to obtain the solution relaxation time $\lambda$ in the shear thickened phase. Irrespective of the {\cds} value, $\lambda$ $\sim$ {\csd}$^{6}$. We defined a Weissenberg number \emph{We} $\equiv \left.{\lambda}\right|_{\psi_{1,max}} \left.{\dot{\gamma}}\right|_{\psi_{1,max}}$, and found that \emph{We} was only weakly dependent on $C_{s}$. This observation was used to develop master curves for viscosity vs. shear rate in terms of suitably reduced variables, which indicated that the SIS formation was self-similar with respect to {\csd}. A qualitative mechanism that combines the effect of {\csd} on average micelle length and Debye parameter with shear-induced configurational changes of rod-like micelles is proposed to rationalize the self-similarity of SIS formation. The mechanistic understanding developed from the experiments can be used in the future to develop a more rigorous mathematic framework and mesoscopic simulations to study flow-induced network formation in dilute micellar solutions.

\section*{ACKNOWLEDGEMENTS}

We gratefully acknowledge NSF grant CBET 0404243 and CBET 0335348 for partial support of this work.

\pagebreak

\begin{figure}[H] \centering \resizebox{0.9\textwidth}{!}{\includegraphics{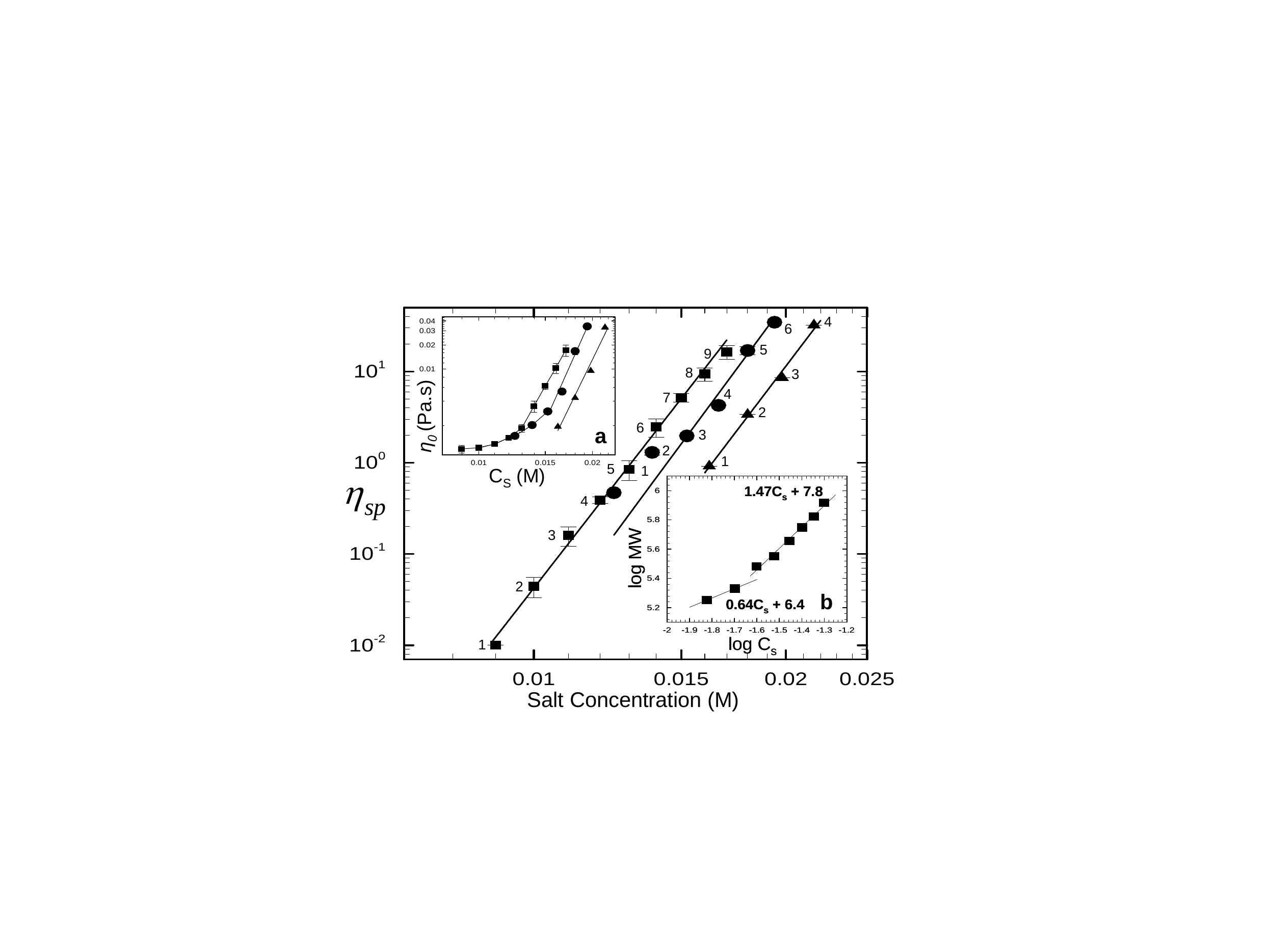}}
\captionsetup{format=capname}
\captionsetup{labelformat=Figurestyle}
\caption{Log-log plot of specific viscosity $\eta_{sp}$ as a function of salt concentration {\csd}, with {\cds} = 0.05M ($\blacksquare$), 0.07M ({\Bullethuge}) and 0.09M ($\blacktriangle$). The numbers 1 $\rightarrow$ 9 denote {\R} = {\csd}/{\cds} values from 0.18 $\rightarrow$ 0.34 respectively in increments of 0.02M. The scaling exponents for the three sets of data range from best power-law fits for the three sets of data are within statistical uncertainty, and range from 11.6 - 12.1. The inset \textbf{a} shows the log-log plot of zero-shear viscosity {\zsv} as a function of salt concentration {\csd}. Lines are only a guide to the eye. The inset \textbf{b} shows the logarithm of molecular weight vs. log {\csd} for {\cds} = 0.05M, and indicates two different slopes signifying a sphere-rod transition.} \label{figure1} \end{figure}
\pagebreak
\begin{figure}[H] \centering \resizebox{0.9\textwidth}{!}{\includegraphics{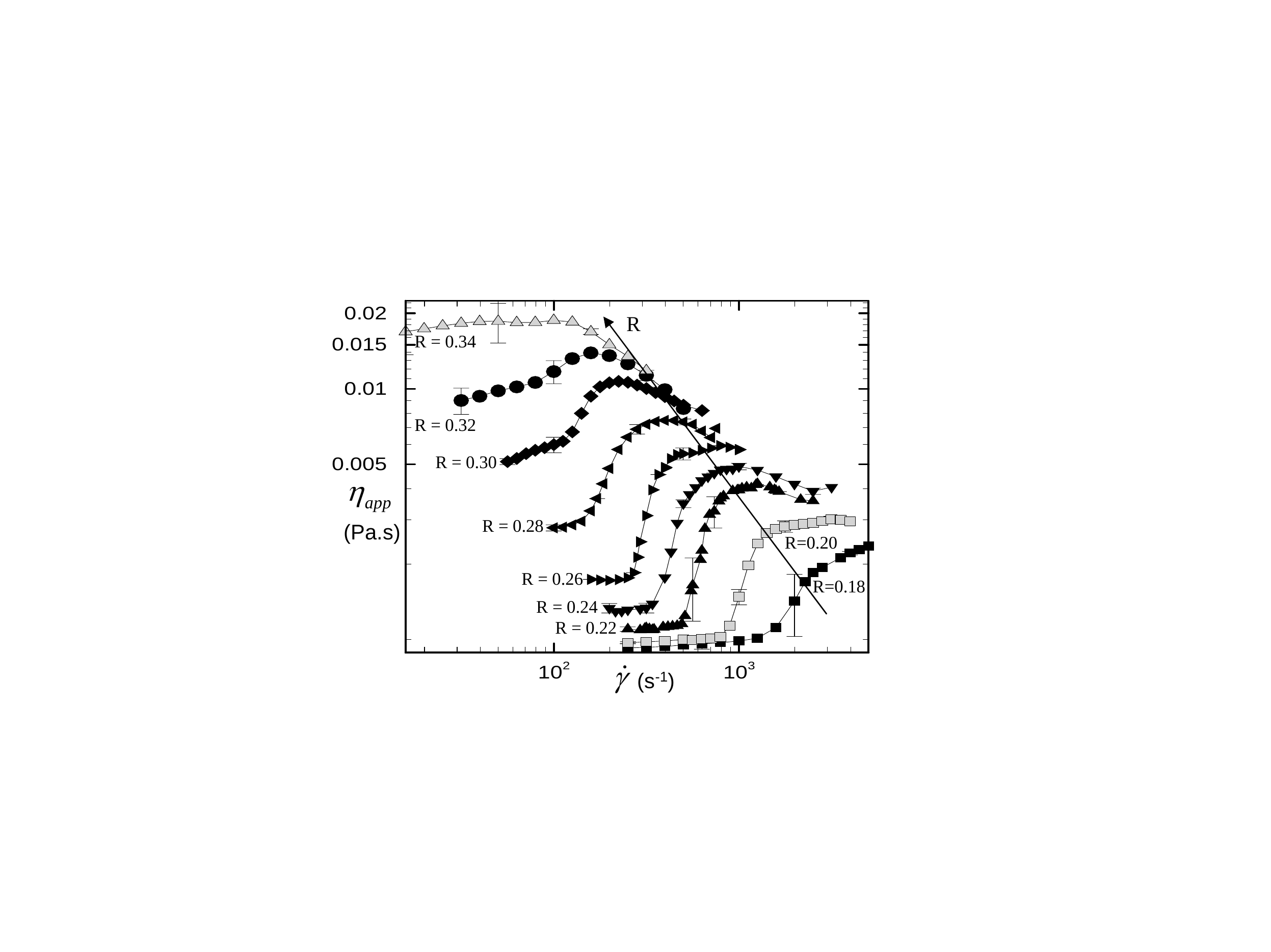}}
\captionsetup{format=capname}
\captionsetup{labelformat=Figurestyle}
\caption{Apparent shear viscosity ($\eta_{app}$) of CTAB $/$ NaSal samples, versus shear rate $\dot{\gamma}$ with R varying from 0.18 to 0.34 in increments of 0.02M. CTAB concentration was fixed at 0.05M. Temperature was maintained at 24$^\circ$C. Lines are only a guide to the eye.} \label{figure2} \end{figure}
\pagebreak
\begin{figure}[H] \centering \resizebox{0.9\textwidth}{!}{\includegraphics{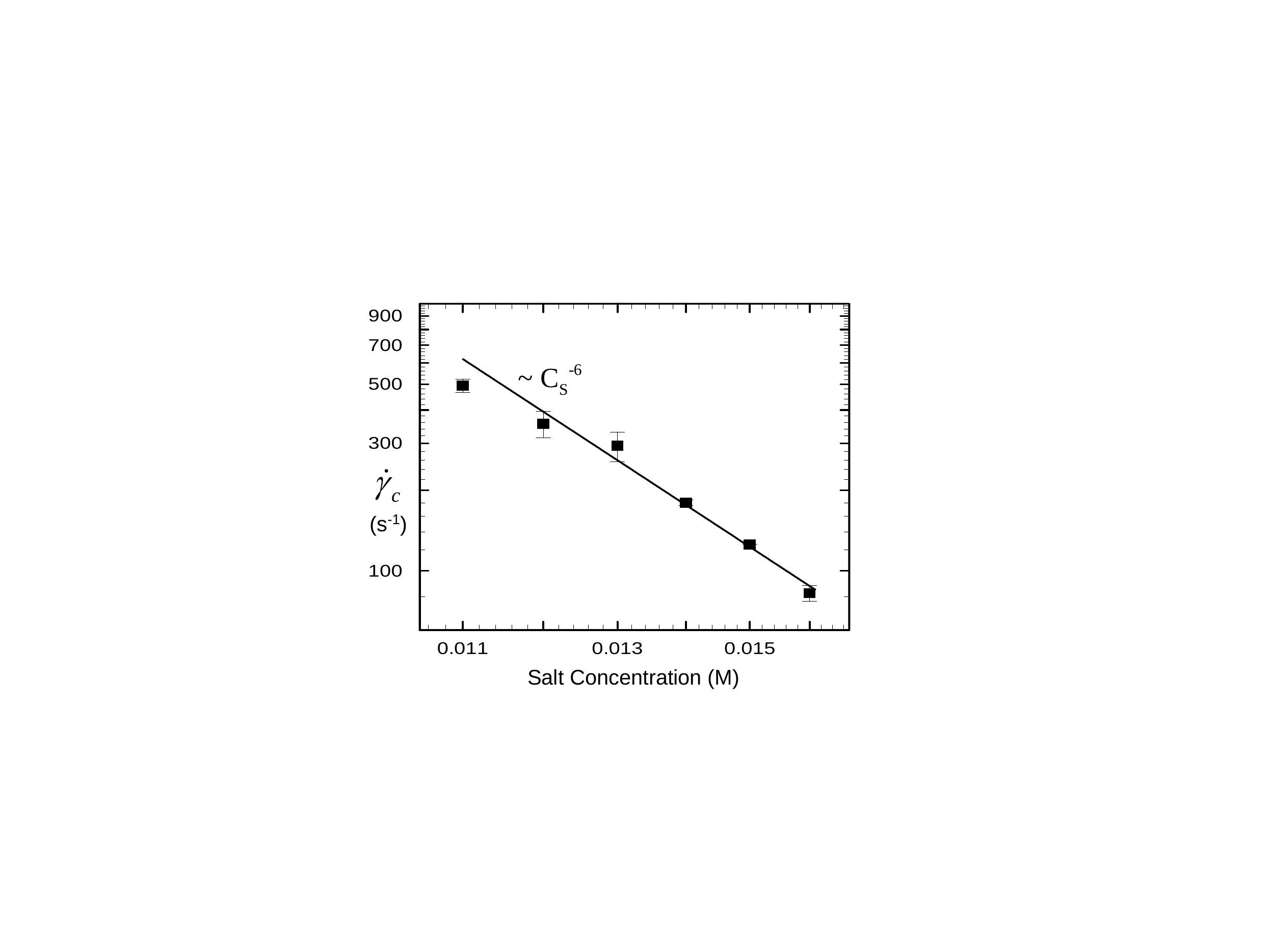}}
\captionsetup{format=capname}
\captionsetup{labelformat=Figurestyle}
\caption{Log-log plot of critical shear rate {\csr} ($\blacksquare$)   as a function of salt concentration {\csd} for solutions with {\cds}=0.05M. The solid line denotes the best power-law fit for {\csr}.} \label{figure3} \end{figure}
\pagebreak
\begin{figure}[H] \centering \resizebox{0.9\textwidth}{!}{\includegraphics{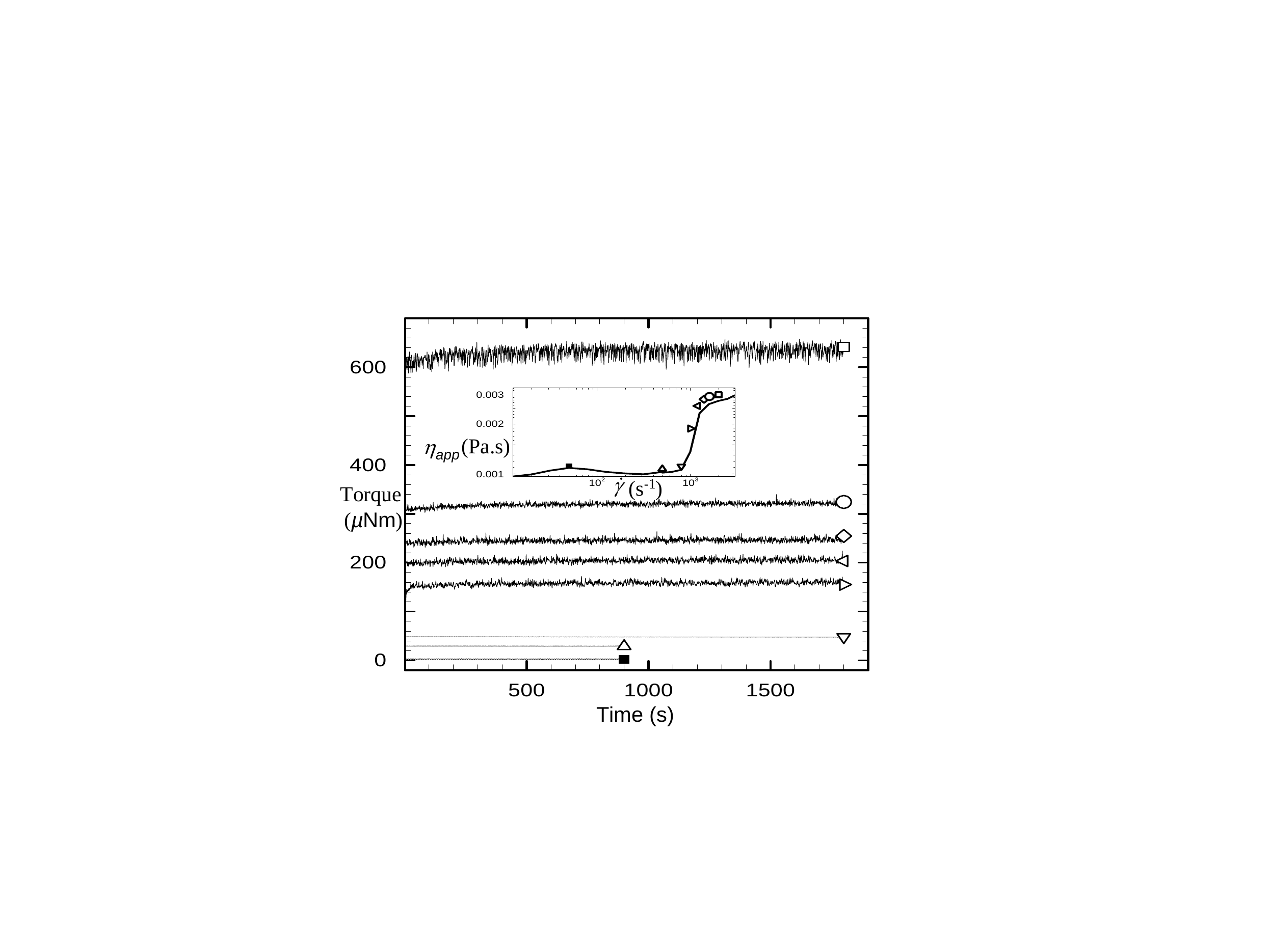}}
\captionsetup{format=capname}
\captionsetup{labelformat=Figurestyle}
\caption{Torque fluctuations vs. time for a solution of ${C_{D}}=0.05M$ and $R=0.2$ at different shear rates. The symbols correspond to $\dot{\gamma}$ $=50$ $({\blacksquare})$, $\dot{\gamma}$ $=500$ $({\vartriangle})$, $\dot{\gamma}$ $=800$ $({\triangledown})$, $\dot{\gamma}$ $=1000$ $({\triangleright})$, $\dot{\gamma}$ $=1200$ $({\triangleleft})$, $\dot{\gamma}$ $=1400$ $({\diamond})$, $\dot{\gamma}$ $=1600$ $({\bigcirc})$, and $\dot{\gamma}$ $=2000$ $({\square})$. Inset$:$ Symbols depict the apparent viscosities averaged over the 30 minute window and the solid line depicts the apparent viscosities averaged for 10 minutes. } \label{figure4} \end{figure}
\pagebreak

\begin{figure}[H] \centering \resizebox{1\textwidth}{!}{\includegraphics{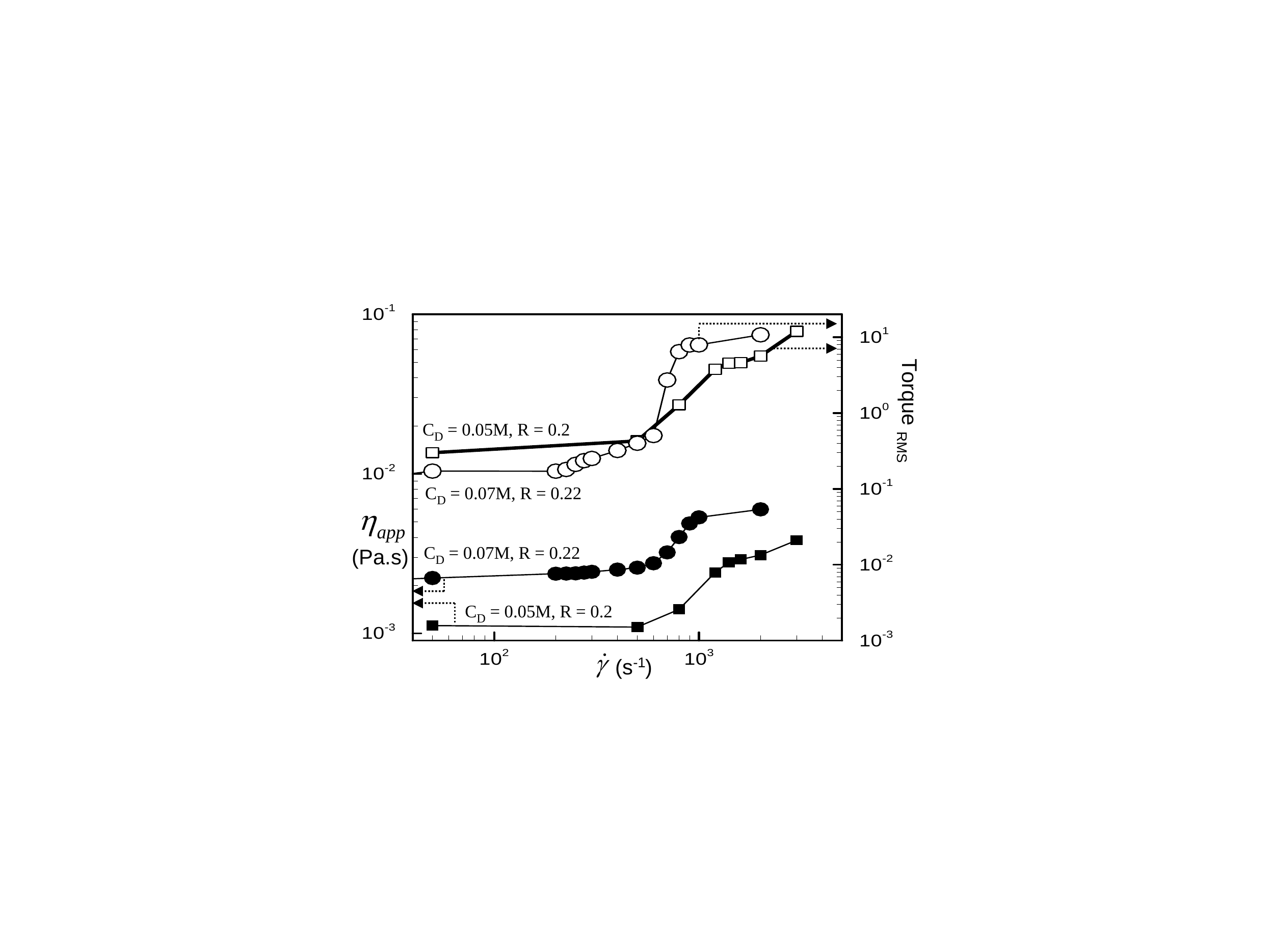}}
\captionsetup{format=capname}
\captionsetup{labelformat=Figurestyle}
\caption{Root Mean Square (R.M.S.) of torque fluctuations (open symbols) shown on the right ordinate, while the apparent viscosity $\eta_{app}$  (filled symbols) is shown on the left ordinate. The symbols ({\Bullethuge},{\circhuge})correspond to solution with {\cds} $=$ 0.07M, {\R} $=$ 0.22, while ($\blacksquare,\square$) correspond to {\cds} $=$ 0.05M, {\R} $=$ 0.2. The viscosity and the torque fluctuations were averaged over 1800 seconds in the shear thickening regime. Lines are only guide to the eye.} \label{figure5} \end{figure}
\pagebreak
\begin{figure}[H] \centering \resizebox{0.9\textwidth}{!}{\includegraphics{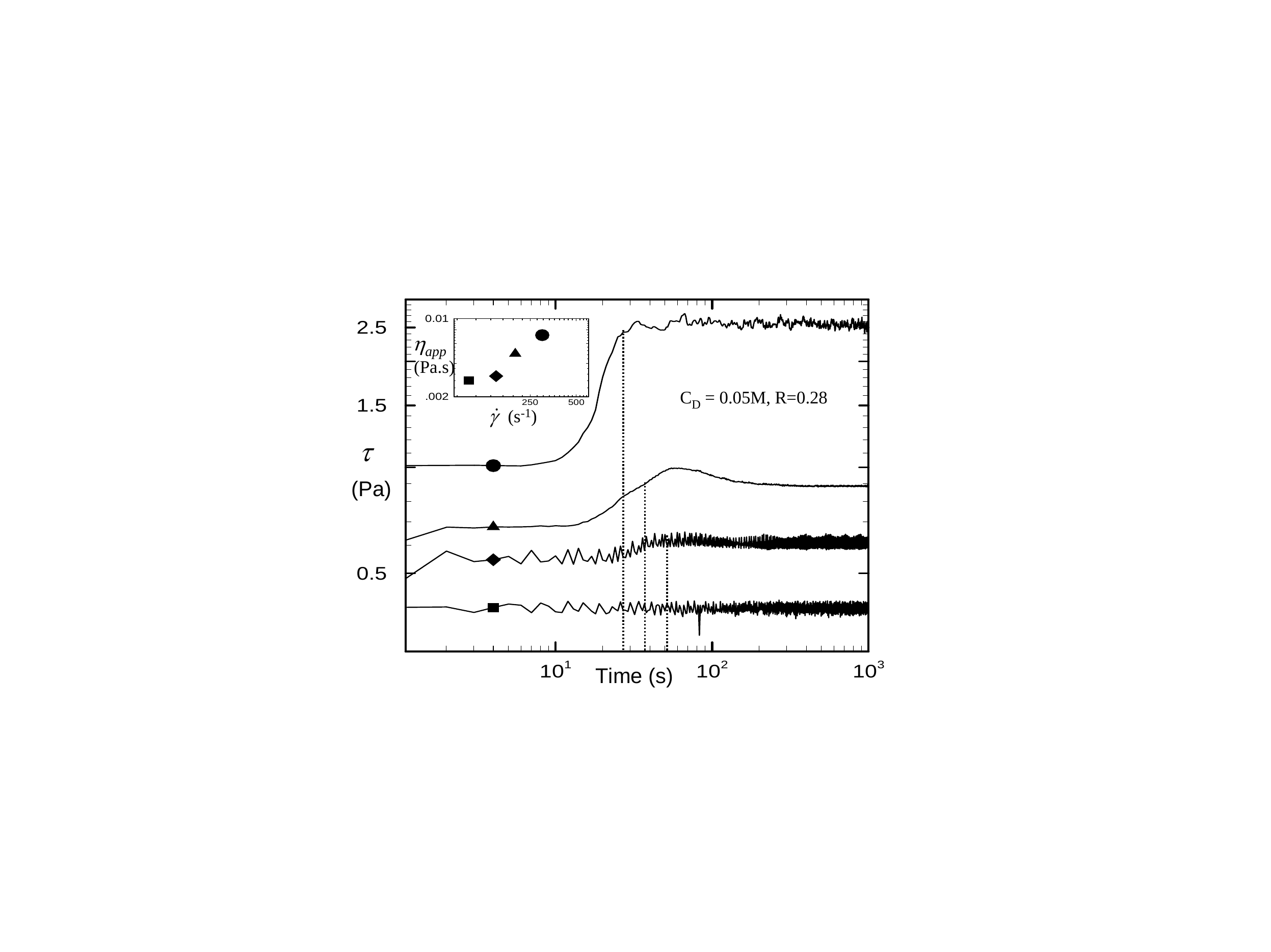}}
\captionsetup{format=capname}
\captionsetup{labelformat=Figurestyle}
\caption{Log-log plot of shear stress  as a function of time for {\sr}= 100 $s^{-1}$ ($\blacksquare$), 150 $s^{-1}$ ($\blacklozenge$), 200 $s^{-1}$ ($\blacktriangle$), and 300 $s^{-1}$ ({\Bullethuge}). Inset shows the {\appvisc} at these shear rates. The time required for stress increase is denoted by the dotted vertical lines.} \label{figure6} \end{figure}
\pagebreak
\begin{figure}[H] \centering \resizebox{0.9\textwidth}{!}{\includegraphics{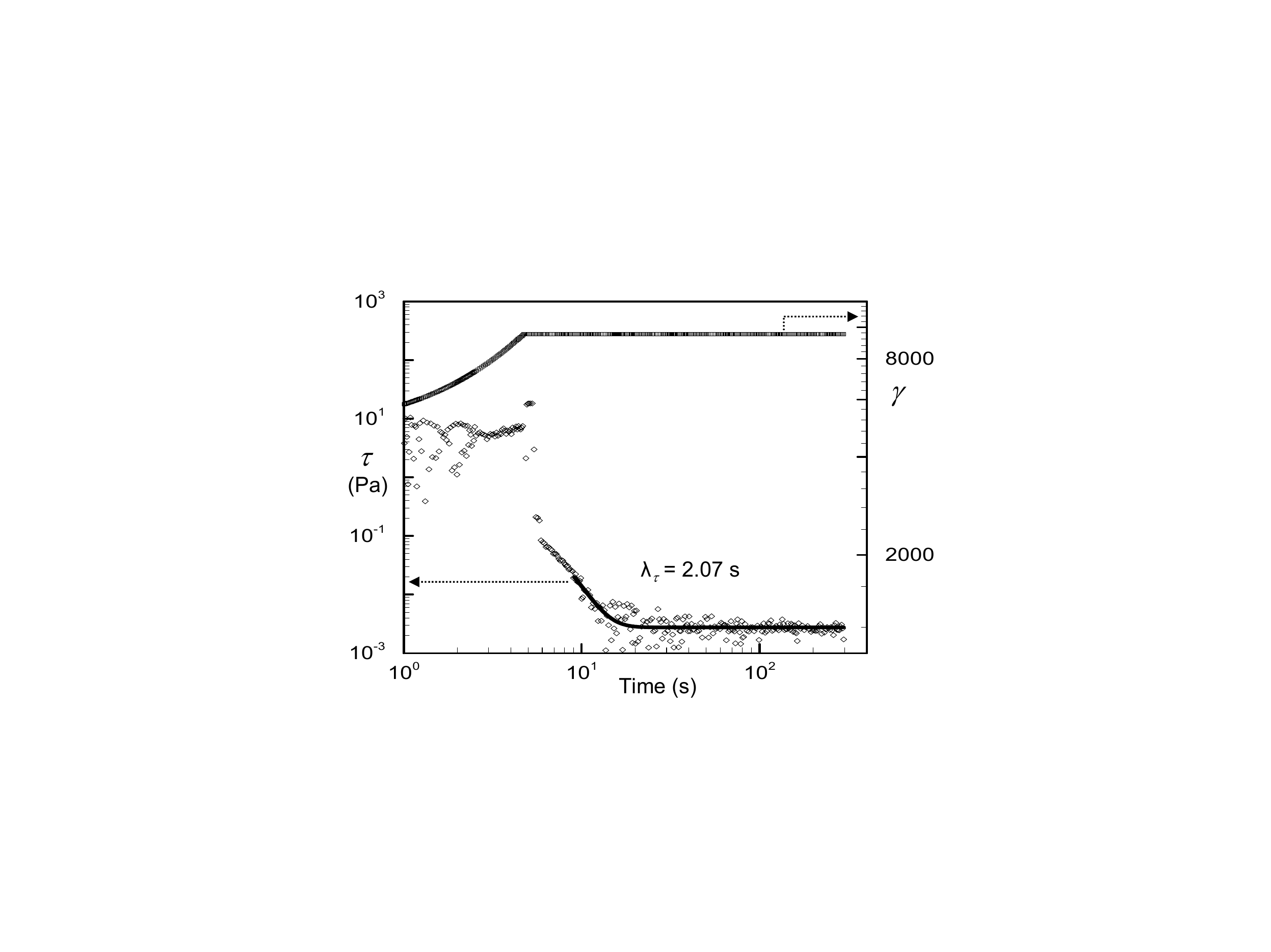}}
\captionsetup{format=capname}
\captionsetup{labelformat=Figurestyle}
\caption{Results of Stress relaxation experiment for {\cds} = 0.05M, {\R} = 0.28. The stress  is displayed as ($\lozenge$) on the left ordinate while the strain ($\gamma$) is displayed as ($\blacksquare$) on the right ordinate. The solid line shows the best exponential fit to the relaxation data.} \label{figure7} \end{figure}
\pagebreak
\begin{figure}[H] \centering \resizebox{0.9\textwidth}{!}{\includegraphics{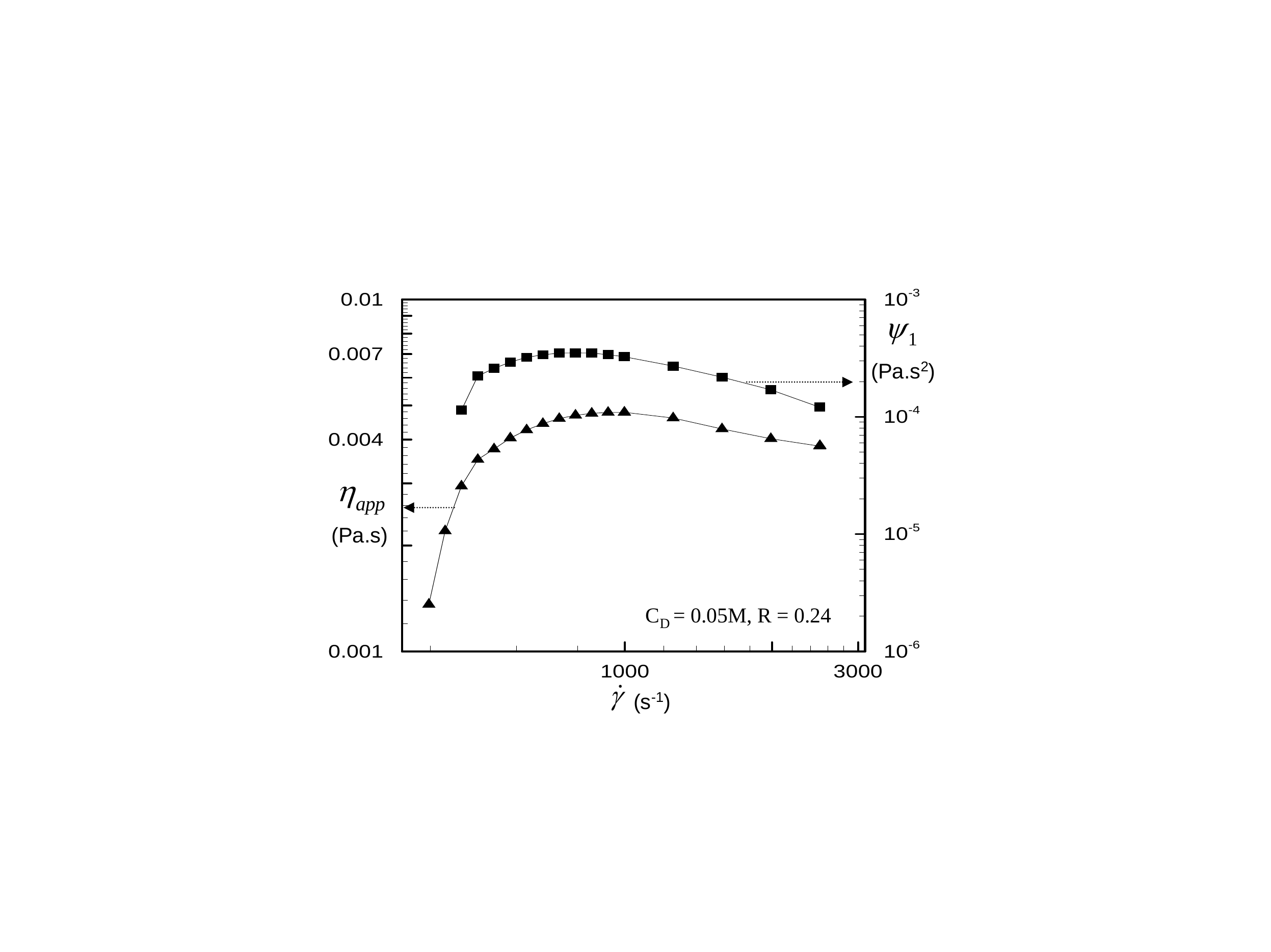}}
\captionsetup{format=capname}
\captionsetup{labelformat=Figurestyle}
\caption{Log-log plot of viscosity $\eta$ ($\blacktriangle$) in the shear-thickening regime  and observed positive first normal stress coefficient $\psi_{1}$ ($\blacksquare$) vs. shear rate {$\dot{\gamma}$} for {\cds}=0.05M and {\R} = 0.24. Lines are only a guide to the eye.} \label{figure8} \end{figure}
\pagebreak
\begin{figure}[H] \centering \resizebox{0.9\textwidth}{!}{\includegraphics{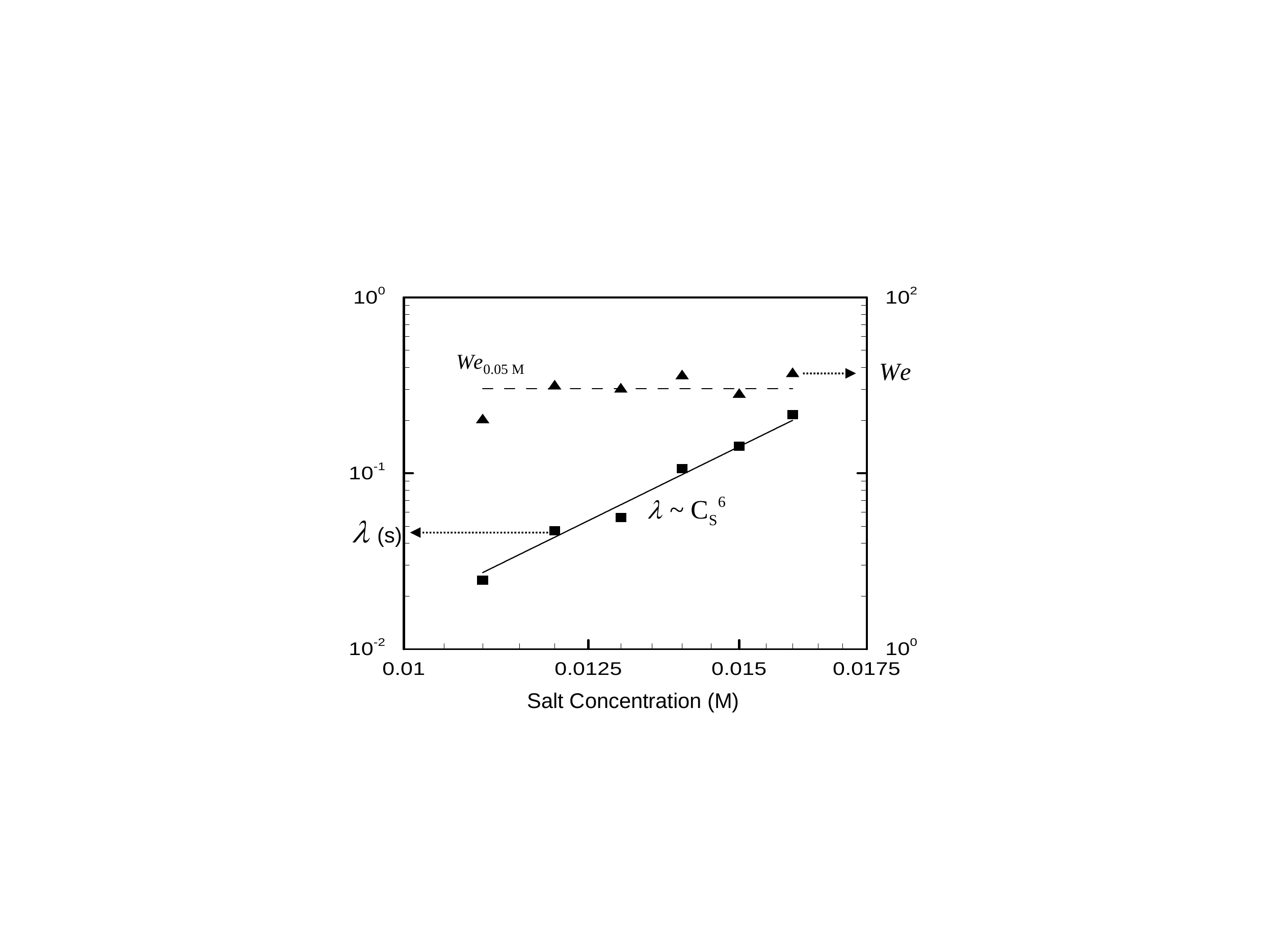}}
\captionsetup{format=capname}
\captionsetup{labelformat=Figurestyle}
\caption{Log-log plot of relaxation time $\lambda$ (left ordinate) and \emph{We} (right  ordinate) as a function of salt concentration for {\cds} = 0.05M. The solid line denotes the best power-law fit for $\lambda$. Dotted line is only a guide to the eye.} \label{figure9} \end{figure}
\pagebreak

\begin{figure}[H] \centering \resizebox{0.9\textwidth}{!}{\includegraphics{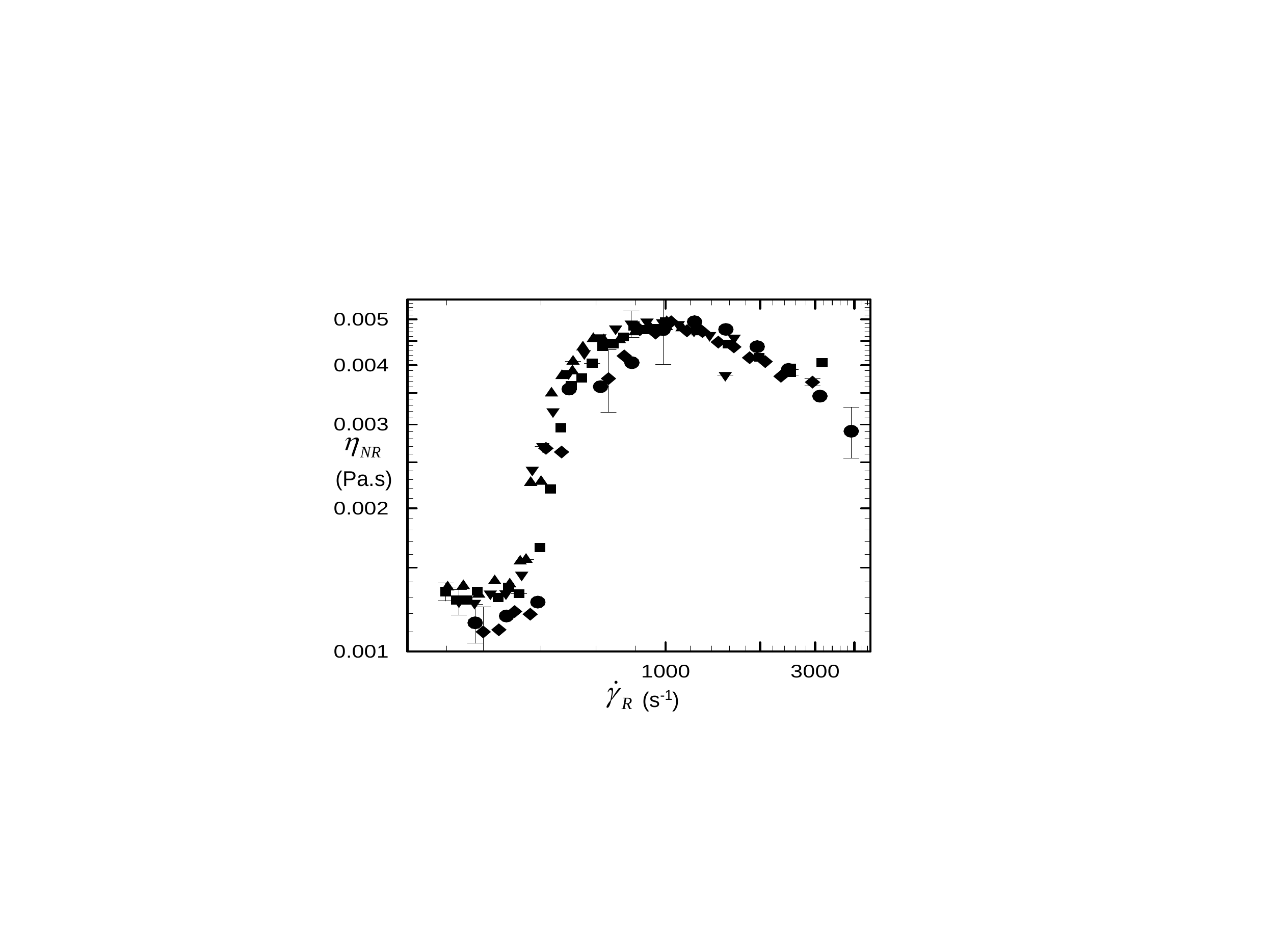}}
\captionsetup{format=capname}
\captionsetup{labelformat=Figurestyle}
\caption{Master curve for reduced normalized viscosity (${\eta_{NR}}$) as a function of reduced shear rate ({$\dot{\gamma_{R}}$}) for solutions with {\cds} = 0.05M and {\R} = 0.24 ($\blacksquare$), 0.26, ($\blacktriangle$), 0.28 ($\blacktriangledown$), 0.30 ($\blacklozenge$) and 0.32 ({\Bullethuge}).} \label{figure10} \end{figure}
\pagebreak
\begin{figure}[H] \centering \resizebox{0.9\textwidth}{!}{\includegraphics{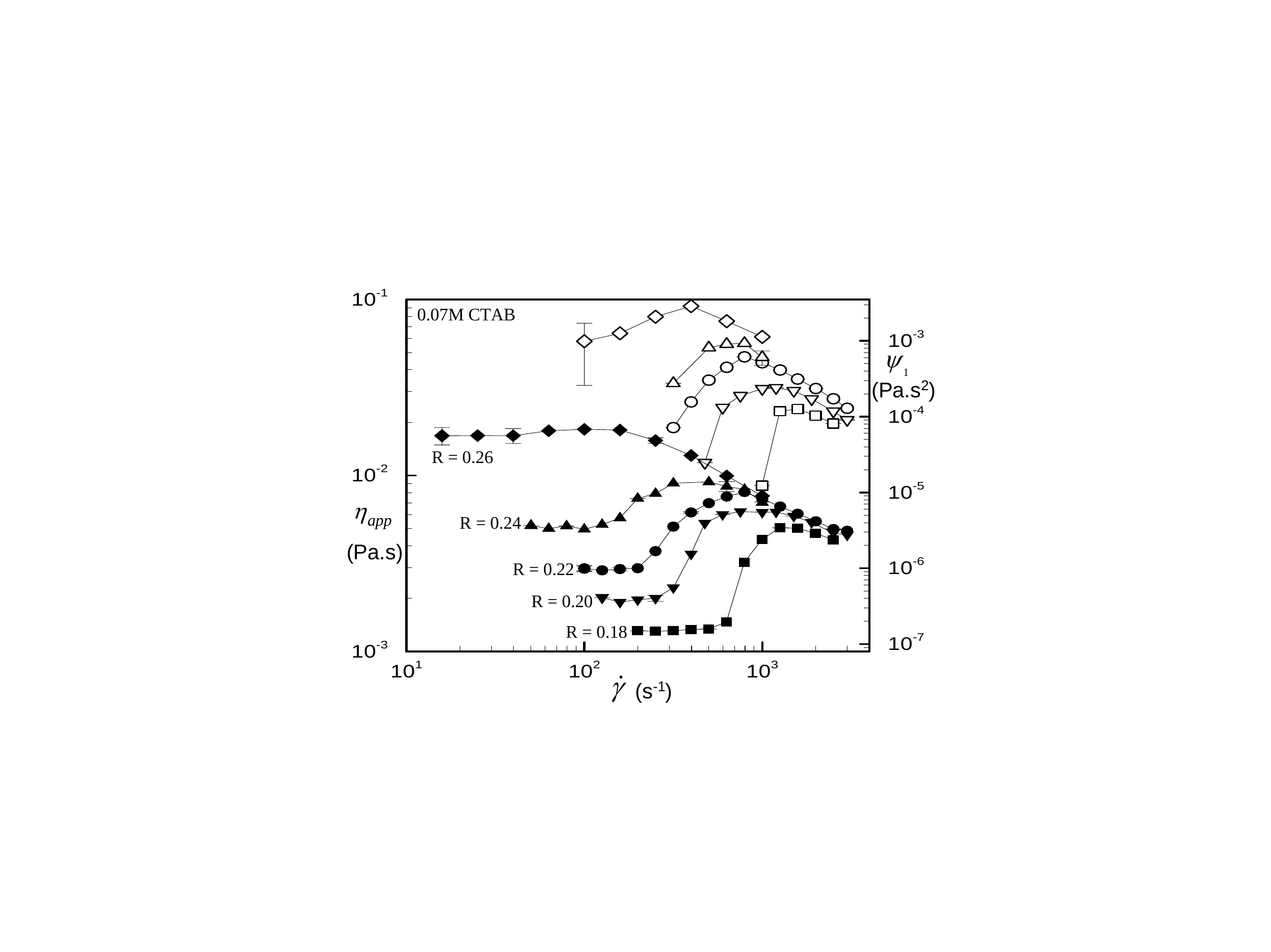}}
\captionsetup{format=capname}
\captionsetup{labelformat=Figurestyle}
\caption{Log-log plot of apparent viscosities {\appvisc} versus shear rate {\sr} is shown on left ordinate  (filled symbols), and the respective first normal stress coefficient $\psi_{1}$ values are shown  on the right ordinate  (open symbols) for {\cds} = 0.07M. The molar ratios are {\R} = 0.18 ($\blacksquare,\square$), {\R} = 0.20 ($\blacktriangledown,\triangledown$), {\R} = 0.22 ({\Bullethuge},{\circhuge}), {\R} = 0.24 ($\blacktriangle,\vartriangle$) and {\R} = 0.26 ($\blacklozenge,\lozenge$).} \label{figure11} \end{figure}
\pagebreak
\begin{figure}[H] \centering \resizebox{0.9\textwidth}{!}{\includegraphics{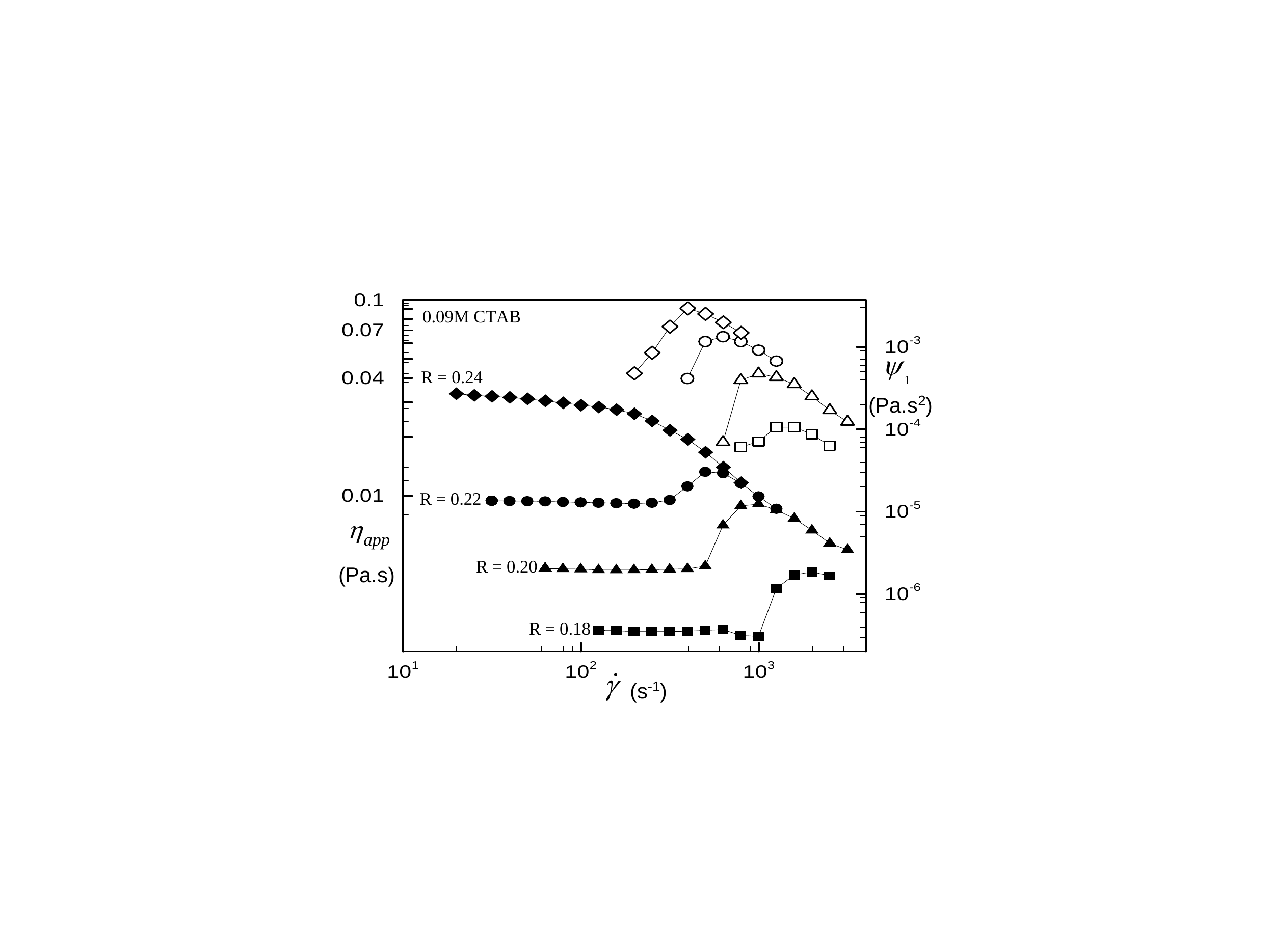}}
\captionsetup{format=capname}
\captionsetup{labelformat=Figurestyle}
\caption{Log-log plot of apparent viscosities {\appvisc} versus shear rate {\sr} is shown on left ordinate  (filled symbols), and the respective first normal stress coefficient $\psi_{1}$ values are shown  on the right ordinate  (open symbols) for {\cds} = 0.09M. The molar ratios are {\R} = 0.18 ($\blacksquare,\square$), {\R} = 0.20 ($\blacktriangledown,\triangledown$), {\R} = 0.22 ({{$\bullet,\circ$}}), and {\R} = 0.24 ($\blacklozenge,\lozenge$).} \label{figure12} \end{figure}
\pagebreak
\begin{figure}[H] \centering \resizebox{0.9\textwidth}{!}{\includegraphics{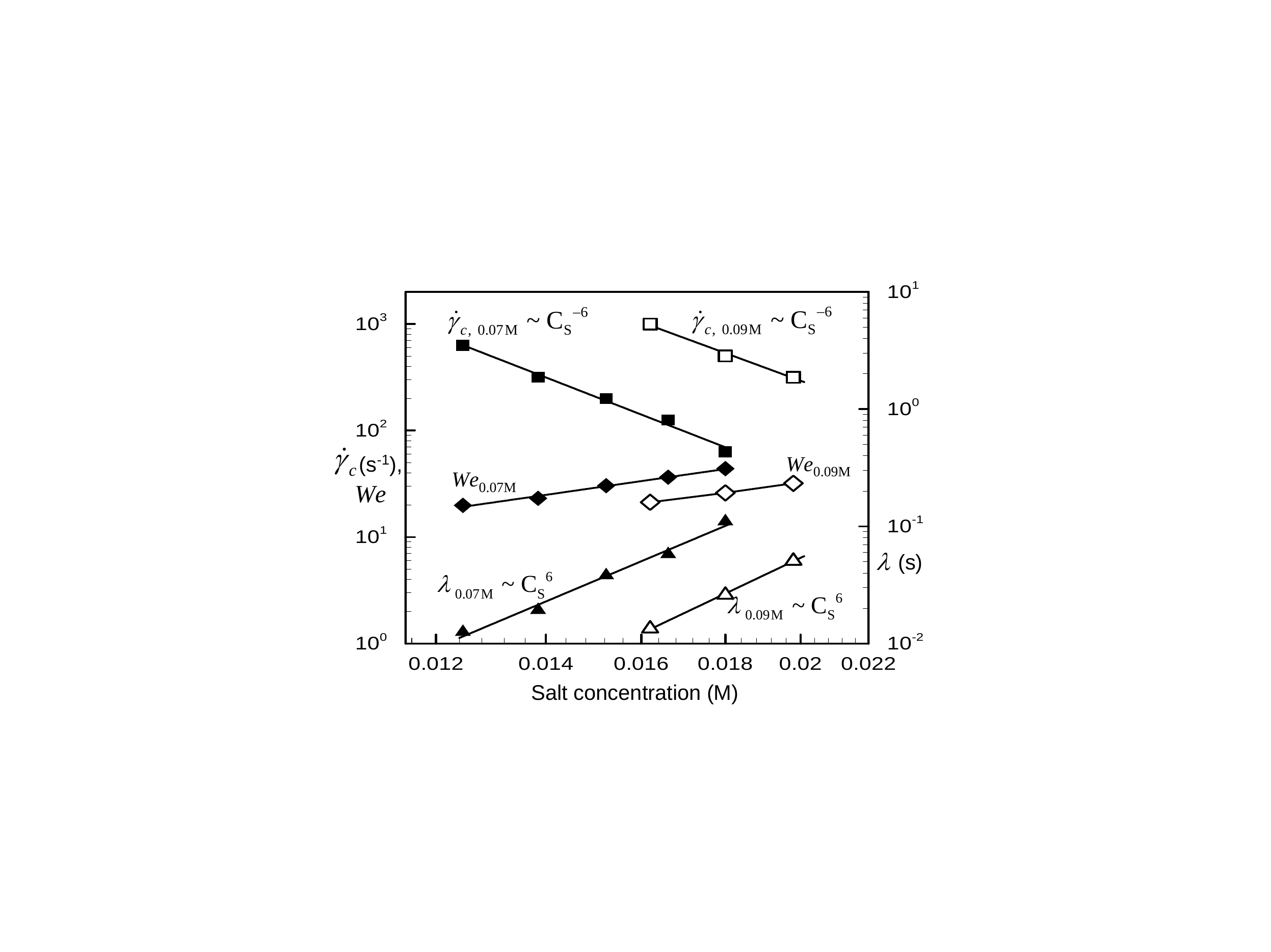}}
\captionsetup{format=capname}
\captionsetup{labelformat=Figurestyle}
\caption{Plot of critical shear rate {\csr} ($\blacksquare$,$\square$), relaxation time $\lambda$ ($\blacktriangle$,$\vartriangle$), and \emph{We} ($\blacklozenge$,$\lozenge$) as a function of salt concentration for {\cds} = 0.07M (filled symbols) and 0.09M (open symbols).} \label{figure13} \end{figure}
\pagebreak
\begin{figure}[H] \centering \resizebox{0.9\textwidth}{!}{\includegraphics{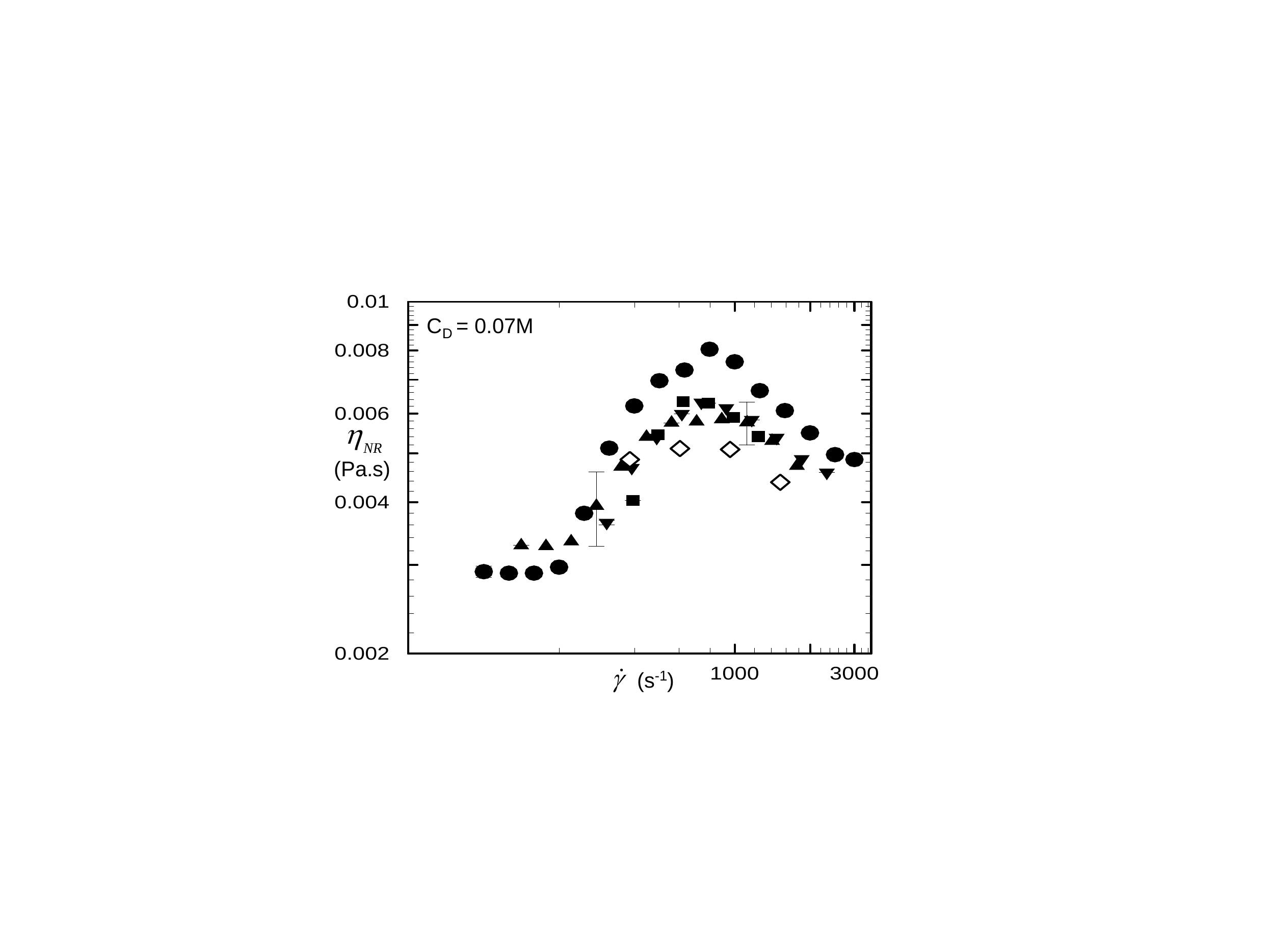}}
\captionsetup{format=capname}
\captionsetup{labelformat=Figurestyle}
\caption{Master curve for ${\eta_{NR}}$ as a function of reduced shear rate {$\dot{\gamma_{R}}$} for solutions with {\cds} = 0.07M and {\R} = 0.18 ($\blacksquare$), 0.20 ($\blacktriangledown$), 0.22 ({\Bullethuge}), 0.24 ($\blacktriangle$) and 0.26 ($\lozenge$).} \label{figure14} \end{figure}
\pagebreak
\begin{figure}[H] \centering \resizebox{0.9\textwidth}{!}{\includegraphics{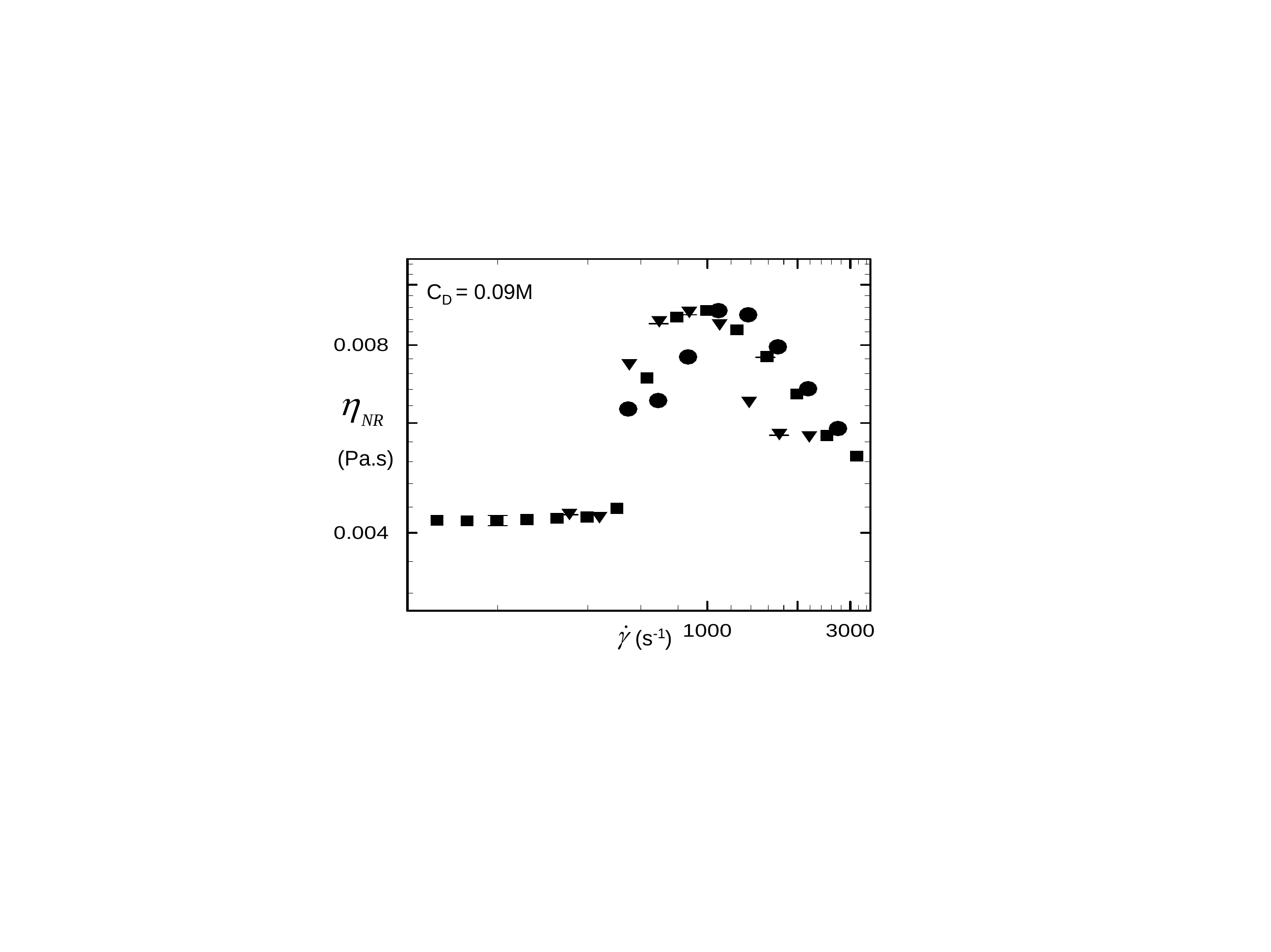}}
\captionsetup{format=capname}
\captionsetup{labelformat=Figurestyle}
\caption{Master curve for ${\eta_{NR}}$ as a function of reduced shear rate {$\dot{\gamma_{R}}$} for solutions with {\cds} = 0.09M and {\R} = 0.18 ($\blacksquare$), 0.20, ($\blacktriangledown$) and 0.22 ({\Bullethuge}).}
\label{figure15} \end{figure}
\pagebreak
\begin{figure} [H] \centering
\resizebox{0.9\textwidth}{!}{\includegraphics{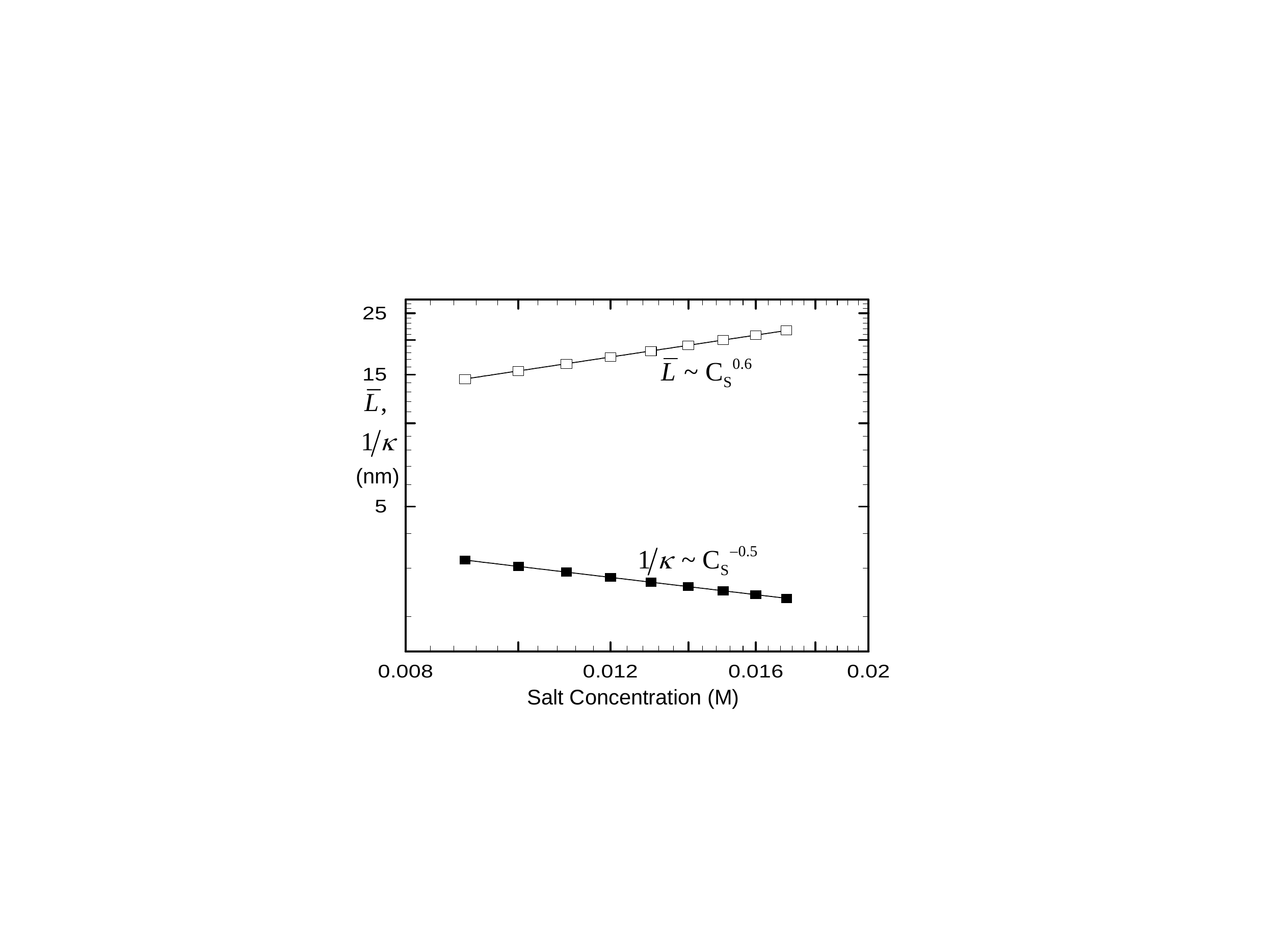}}
\captionsetup{format=capname}
\captionsetup{labelformat=Figurestyle}
\caption{Log-log plot of average micelle length $\bar{L}$ ($\square$) and Debye length 1/$\kappa$ ($\blacksquare$) as a function of salt concentration for solutions with {\cds} = 0.05M.} \label{figure16} \end{figure}
\pagebreak
\begin{figure} [H] \centering
\resizebox{0.9\textwidth}{!}{\includegraphics{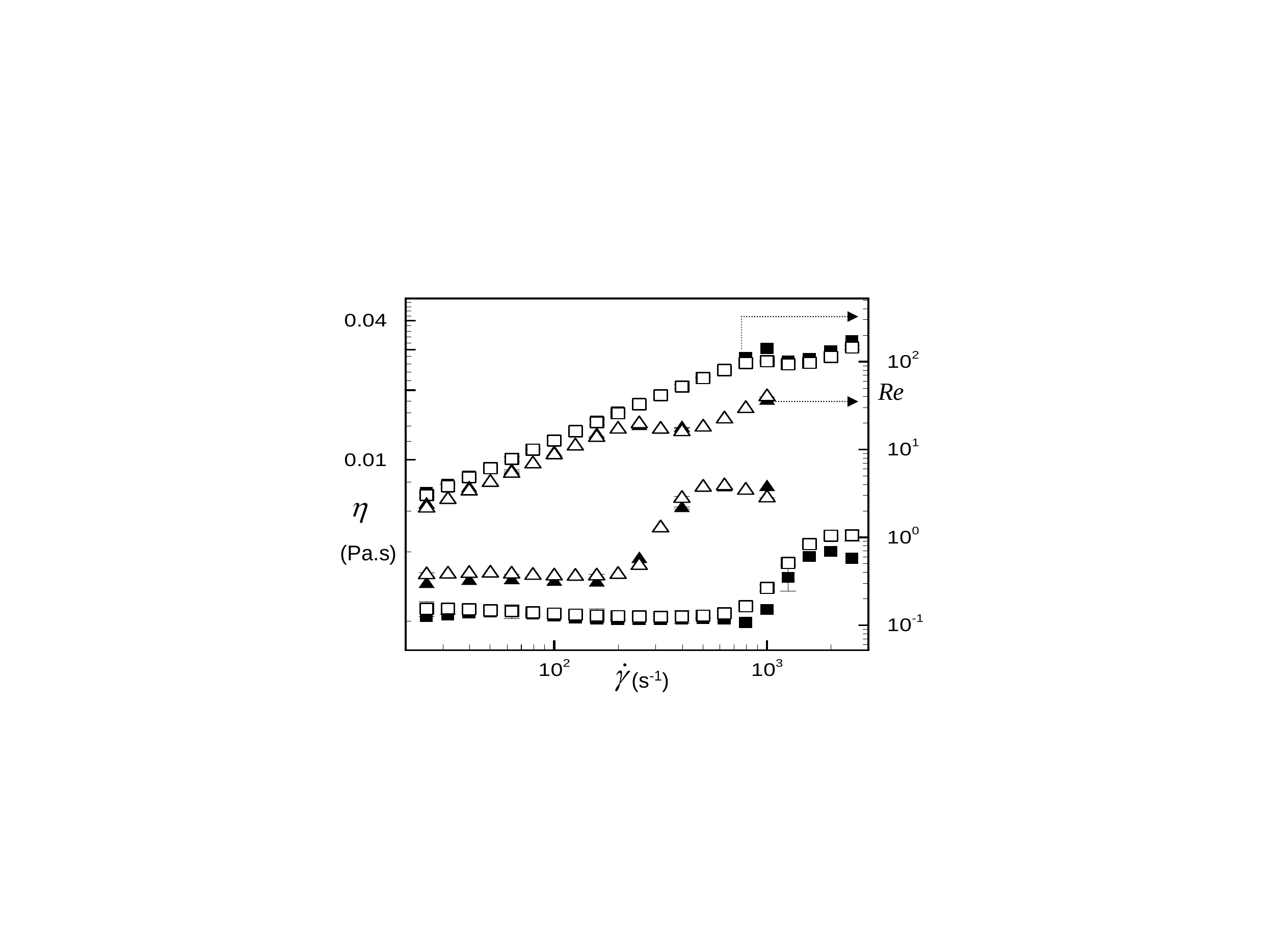}}
\captionsetup{format=capname}
\captionsetup{labelformat=Figurestyle}
\caption{Comparison of {\appvisc} values obtained from Cone-and-Plate (filled symbols) and Taylor-Couette (open symbols) for solutions with {\cds} = 0.05M, {\R} = 0.2 ($\square, \blacksquare$) and {\cds} = 0.07M, {\R} = 0.2 ($\vartriangle, \blacktriangle$). The respective Reynolds numbers {\it{Re}} are shown on the right ordinate.} \label{figure17} \end{figure}
\end{document}